\documentclass[12pt,a4paper,american]{article}
\usepackage[T1]{fontenc}
\usepackage[latin9]{inputenc}
\usepackage{amsmath}
\usepackage{graphicx}
\usepackage{amssymb}
\usepackage{esint}

\makeatletter
\newcommand{\lyxaddress}[1]{
\par {\raggedright #1
\vspace{1.4em}
\noindent\par}
}

\usepackage{mathrsfs}
\usepackage{amsthm}

\renewcommand{\Re}{\mathop\mathrm{Re}\nolimits}

\newcommand{\Dom}{\mathop\mathrm{Dom}\nolimits}
\newcommand{\Ran}{\mathop\mathrm{Ran}\nolimits}
\newcommand{\Ker}{\mathop\mathrm{Ker}\nolimits}
\newcommand{\Res}{\mathop\mathrm{Res}\nolimits}

\newtheorem{theorem}{Theorem}
\newtheorem{proposition}[theorem]{Proposition}
\newtheorem{lemma}[theorem]{Lemma}

\theoremstyle{remark}
\newtheorem{remark}[theorem]{Remark}

\newcommand{\sH}{\mathscr{H}}
\newcommand{\sT}{\mathscr{T}}

\makeatother

\usepackage{babel}

\makeatother

\usepackage{babel}

\begin{document}

\title{On the two-dimensional Coulomb-like potential with a central point
interaction}

\date{{}}

\author{P.~Duclos$^{1}$%
\thanks{Pierre Duclos passed away in January 2010. The manuscript was prepared
for publication by the second and the third author.%
}~, P.~\v{S}\v{t}ov\'\i\v{c}ek$^{2}$, M.~Tu\v{s}ek$^{2}$}

\maketitle

\lyxaddress{$^{1}$Centre de Physique Th\'eorique (CPT-CNRS UMR 6207) Universit\'e
du Sud, Toulon-Var, BP 20132, F--83957 La Garde Cedex, France}

\lyxaddress{$^{2}$Department of Mathematics, Faculty of Nuclear Science, Czech
Technical University in~Prague, Trojanova 13, 120~00 Praha, Czech
Republic}
\begin{abstract}
\noindent In the first part of the paper, we introduce the Hamiltonian
$-\Delta-Z/\sqrt{x^{2}+y^{2}}$, $Z>0$, as a self-adjoint operator
in $L^{2}(\mathbb{R}^{2})$. A general central point interaction combined
with the two-dimensional Coulomb-like potential is constructed and
properties of the resulting one-parameter family of Hamiltonians is
studied in detail. The construction is also reformulated in the momentum
representation and a relation between the coordinate and the momentum
representation is derived. In the second part of the paper we prove
that the two-dimensional Coulomb-like Hamiltonian can be derived as
a norm resolvent limit of the Hamiltonian of a Hydrogen atom in a
planar slab as the width of the slab tends to zero.\\

\noindent \vspace{\baselineskip}\emph{ PACS}: 02.30.Sa, 02.30.Tb,
03.65.Db
\end{abstract}

\section{Introduction}

In this paper we discuss, in the framework of nonrelativistic quantum
mechanics, two subjects related to the two-dimensional Coulomb-like
potential in the plane. In the first part, in Section~\ref{sec:point},
we reexamine the two-dimensional Hydrogen atom. This is to say that
we consider a quantum model in the plane with the attractive potential
\begin{equation}
V(x,y)=-Z/\varrho,\ \varrho=\sqrt{x^{2}+y^{2}}\,,\label{eq:2D_Coul}\end{equation}
which we call the two-dimensional Coulomb (or hydrogenic) potential.
This model has already been studied from various points of view in
the physical literature. A detailed analysis of this system is given
in \cite{ygc}, see also \cite{pp}. The corresponding Green function
is constructed in \cite{hostler}, though with some minor misprints.
More mathematically oriented questions like the proper definition
of the Hamiltonian as a self-adjoint operator are shortly discussed
in the recent paper \cite{dov}. Note that some interest to this type
of models also comes from semiconductor physics. In a semiconductor
quantum well under illumination, excited electrons and holes are essentially
confined to the plane and interact via a mutual Coulomb interaction
which results in the creation of electron-hole bound states, known
as excitons \cite{pp}. Moreover, as shown in the second part of the
paper, the two-dimensional hydrogenic Hamiltonian can be viewed as
an approximation of the Hamiltonian of a Hydrogen atom in a thin planar
layer.

Let us point out, however, that if a Hydrogen atom is supposed to
be two-dimensional in the strict sense, i.e. all fields including
electromagnetic fields, the angular momentum, and the spin are confined
to the plane, then (\ref{eq:2D_Coul}) is no longer eligible to as
the two-dimensional Coulomb potential. Indeed, the Coulomb law may
be derived from the first Maxwell equation (Gauss's law for electrostatics)
stating that $\mathrm{div}\,\mathbf{E}=\sigma$ where $\sigma$ stands
for the planar charge density, $E_{z}=0$, and the electric field
is supposed to be rotationally symmetric. Integration of this equation
over a disk of radius $\varrho$ together with application of Green's
theorem leads to the choice of the potential in the form \[
V(x,y)=\mbox{const}\,\ln(\varrho).\]
The Schr\"odinger equation for this potential is studied in \cite{adl}.

One of the goals of the present paper is to describe a central point
interaction combined with the two-dimensional Coulomb-like potential
and to study its basic properties. The construction of point interactions
based on the theory of self-adjoint extensions is now pretty well
established. To our best knowledge, however, the two-dimensional Coulomb-like
potential is not yet discussed in the literature, including the well
known monograph \cite{ag} where only the one-dimensional and the
three-dimensional cases are considered. On the other hand, there exists
a general theoretical background for the construction of self-adjoint
extensions with singular boundary conditions, as described in paper
\cite{bulla_ges}, which is directly applicable to our model.

Along with the construction of point interactions in the coordinate
representation, and this is the standard way how to proceed, we discuss
the construction also in the momentum representation. Moreover, we
derive an explicit relation between the two representations. This
correspondence is based on the Whittaker integral transformation whose
integral kernel is a properly normalized generalized eigenfunction
of the two-dimensional Coulomb-like Hamiltonian depending on the spectral
parameter. Remarkably, this integral transformation has been studied
in the mathematical literature quite recently \cite{AlMusallamTuan,Fulton}.
On this point we refer, first of all, to paper \cite{GesztesyZinchenko}
where the unitarity of the eigenfunction expansion is proven for a
much more general class of Schr\"odinger operators on a halfline.

In the second part of the present paper, in Section~\ref{sec:layer},
we study the Hydrogen atom in a thin planar layer of width $a$, called
$\Omega_{a}$. In our model, we confine the atom to the slab by imposing
the Dirichlet boundary condition on the parallel boundary planes.
Our main goal in this section is to show that the resulting Hamiltonian
in $L^{2}(\Omega_{a})$, called $H^{a}$, is well approximated in
a convenient sense by the two-dimensional Coulomb-like Hamiltonian
as the width of the layer approaches zero. The method we use is strongly
motivated by the paper by Brummelhuis and Duclos \cite{bd}. Firstly,
we apply the projection on the first transversal mode getting this
way the so-called effective Hamiltonian in $L^{2}(\mathbb{R}^{2})$.
Then, in Subsection~\ref{sec:coul_eff}, we show that the norm resolvent
limit of the effective Hamiltonian, as $a\to0+$, exactly equals the
two-dimensional hydrogenic Hamiltonian plus the energy of the lowest
transversal mode. As a next step we prove, in Subsection~\ref{sec:eff_aprox},
that the full Hamiltonian $H^{a}$ is well approximated by the effective
Hamiltonian, again in the norm resolvent sense. Since the spectrum
of $H_{C}$ is known explicitly one can use this approximation to
derive, with the aid of standard perturbation methods, asymptotic
formulas for the eigenvalues of $H^{a}$ though we do not go into
details at this point.

\section{Two-dimensional Coulomb-like potential with a central point interaction
\label{sec:point}}

\subsection{The coordinate representation}

Let $-\Delta$ be the free Hamiltonian in $L^{2}(\mathbb{R}^{2},\mbox{d}x\mbox{d}y)$.
It is known that the Coulomb-like potential $\left(x^{2}+y^{2}\right)^{\!-1/2}$
in the plane is $(-\Delta)$ form bounded with relative bound zero.
This is a consequence of the Kato inequality; the proof is given in
\cite{bouzouina} but see also \cite{herbst} where even a more general
case is treated. In more detail, the inequality claims that \begin{equation}
\frac{1}{\sqrt{x^{2}+y^{2}}}\,\leq\,\frac{\Gamma(1/4)^{4}}{4\pi^{2}}\,\sqrt{-\Delta}\,.\label{eq:Kato_ineq}\end{equation}
Suppose $Z>0$. By the KLMN theorem \cite[Theorem~X.17]{rs2}, the
operator\begin{equation}
H_{C}=-\Delta-\frac{Z}{\sqrt{x^{2}+y^{2}}}\ \mbox{\ (the form sum)}\label{eq:H_C_def}\end{equation}
is self-adjoint. The form domain of the Hamiltonian $H_{C}$ coincides
with that of the free Hamiltonian, i.e. with the first Sobolev space.
In particular, $\Dom(H_{C})\subset\mathcal{H}^{1}(\mathbb{R}^{2})$.
Note that the same conclusions can also be deduced from the results
in \cite[Chp.~XIII.11]{rs4}. The operator $H_{C}$ has been studied
quite intensively in the physical literature (see, for example, \cite{hostler,ygc,pp}).

To introduce a central point interaction let us consider the densely
defined symmetric operator,\[
\dot{H}=-\Delta-\frac{Z}{\sqrt{x^{2}+y^{2}}}\,,\ \Dom(\dot{H})=C_{0}^{\infty}(\mathbb{R}^{2}\setminus\left\lbrace 0\right\rbrace ).\]
Denote by $H_{\mathrm{min}}$ the closure of $\dot{H}$. Then $H_{C}$
is exactly the Friedrichs extension of $H_{\mathrm{min}}$. As usual,
the Hilbert space naturally decomposes in the polar coordinates $(\varrho,\varphi)$,\[
L^{2}(\mathbb{R}^{2},\mbox{d}x\mbox{d}y)=\bigoplus_{m=-\infty}^{\infty}L^{2}(\mathbb{R}_{+},\varrho\,\mbox{d}\varrho)\otimes\mathbb{C}e^{im\varphi},\]
and the operator $H_{\mathrm{min}}$ decomposes correspondingly,\[
H_{\mathrm{min}}=\bigoplus_{m=-\infty}^{\infty}H_{\mathrm{min},m}\otimes1,\]
where $H_{\mathrm{min},m}$ is the closure of the operator\[
\dot{H}_{m}=-\frac{\partial^{2}}{\partial\varrho^{2}}-\frac{1}{\varrho}\frac{\partial}{\partial\varrho}+\frac{m^{2}}{\varrho^{2}}-\frac{Z}{\varrho}\,,\ \Dom(\dot{H}_{m})=C_{0}^{\infty}(\mathbb{R}_{+}).\]
Put also $H_{\mathrm{max},m}=H_{\mathrm{min,}m}^{\,\dagger}$. For
the maximal operator one has \cite[Chapter~8]{weidmann} \begin{align*}
 & \Dom(H_{\mathrm{max},m})=\left\lbrace f\in L^{2}(\mathbb{R}_{+},\varrho\,\mathrm{d}\varrho);~f,f'\in AC_{\text{loc}}(\mathbb{R}_{+}),\, L_{m}f\in L^{2}(\mathbb{R}_{+},\mathrm{\varrho\, d}\varrho)\right\rbrace ,\\
 & \mbox{with\ }L_{m}=-\frac{\partial^{2}}{\partial\varrho^{2}}-\frac{1}{\varrho}\frac{\partial}{\partial\varrho}+\frac{m^{2}}{\varrho^{2}}-\frac{Z}{\varrho}\,.\end{align*}
If $f\in\Dom(H_{\mathrm{max},m})$ then $H_{\mathrm{max},m}f=L_{m}f$.

For $z\in\mathbb{C}\setminus\mathbb{R}$ and $m\in\mathbb{Z}$ consider
the equation $(L_{m}-z)f=0$. Two independent solutions are expressible
in terms of the Whittaker functions, namely\begin{equation}
\varrho^{-1/2}M_{Z/(2\sqrt{-z}),|m|}\!\left(2\sqrt{-z}\,\varrho\right)\mbox{\ and\,\ }\varrho^{-1/2}W_{Z/(2\sqrt{-z}),|m|}\!\left(2\sqrt{-z}\,\varrho\right)\!,\label{eq:WhittakerMW}\end{equation}
with $\Re\sqrt{-z}>0$. From the asymptotic expansions it follows
(see, for instance, \cite{as}) that the former function in (\ref{eq:WhittakerMW})
is square integrable at zero but is not square integrable at infinity
while the latter one is square integrable at infinity but is not square
integrable at zero, except of the case $m=0$. Thus for $m\neq0$,
the operators $H_{\mathrm{min},m}=H_{\mathrm{max},m}=H_{m}$ are self-adjoint
while for $m=0$, $H_{\mathrm{min},0}$ has deficiency indices $(1,1)$.
For a wide class of Schr\"odinger operators, including our case as
well, an explicit construction of all self-adjoint extensions defined
by boundary conditions can be found in \cite{bulla_ges}.

\begin{proposition} All self-adjoint extensions of $H_{\mathrm{min},0}$
are $H_{0}(\kappa)\subset H_{\mathrm{max},0}$, $\kappa\in\mathbb{R}\cup\{\infty\}$,
with the domains\[
\Dom\left(H_{0}(\kappa)\right)=\left\lbrace f\in\Dom(H_{\mathrm{max},0});\, f_{1}=\kappa f_{0}\right\rbrace ,\]
 where the boundary values $f_{0},\, f_{1}$ are defined by \begin{equation}
f_{0}=\lim_{\varrho\to0+}(-\ln\varrho)^{-1}f(\varrho),\ f_{1}=\lim_{\varrho\to0+}\left(f(\varrho)+f_{0}\ln\varrho\right).\label{eq:bvalues}\end{equation}
The self-adjoint extension $H_{0}(\infty)$ determined by the boundary
condition $f_{0}=0$ coincides with the Friedrichs extension of $H_{\mathrm{min},0}$.
\end{proposition}

All self-adjoint extensions $H(\kappa)$ of $H_{\mathbb{\mathrm{min}}}$
are again labeled by $\kappa\in\mathbb{R}\cup\{\infty\}$ and are
equal to\[
H(\kappa)=\bigoplus_{m=-\infty}^{-1}H_{m}\oplus H_{0}(\kappa)\oplus\bigoplus_{m=1}^{\infty}H_{m}.\]
In particular, $H(\infty)$ coincides with $H_{C}$.

\begin{proposition} For the essential spectrum one has $\sigma_{\mathrm{ess}}(H_{C})=[\,0,\infty)$
and, more generally, $\sigma_{\mathrm{ess}}(H(\kappa))=[\,0,\infty)$
for all $\kappa\in\mathbb{R}\cup\{\infty\}$. \end{proposition}

\begin{proof} Let us introduce (temporarily) the functions\[
U_{1}(x,y)=-\frac{Z}{\sqrt{x^{2}+y^{2}+1}}\,,\ U(x,y)=-\frac{Z}{\sqrt{x^{2}+y^{2}}}+\frac{Z}{\sqrt{x^{2}+y^{2}+1}}\,,\]
and denote by $U_{1}$ and $U$ the corresponding multiplication operators.
Put $A=-\Delta+U_{1}$. Since $U_{1}(x,y)$ is bounded and tends to
zero at infinity one knows that $\sigma_{\mathrm{ess}}(A)=[\,0,\infty)$
(see, for instance, \cite[Theorem~4.1]{BerezinShubin}). Note that,
by the closed graph theorem, $(A+k)^{-1/2}(-\Delta+1)^{1/2}$ is bounded
for $k>0$ sufficiently large. Moreover, $U$ is a relatively form
bounded perturbation of $A$ and $H_{C}$ equals the form sum $A+U$.
It is shown below, in the proof of Lemma~\ref{lem:gen_dif_est},
that the operator $(-\Delta+1)^{-1/2}U(-\Delta+1)^{-1/2}$ is Hilbert-Schmidt
and hence compact. Consequently, $U$ is a relatively form-compact
perturbation of $A$ and, by the results in \cite[Chp.~XIII.4]{rs4}
related to the Weyl theorem, $\sigma_{\mathrm{ess}}(H_{C})=\sigma_{\mathrm{ess}}(A)$.
To extend the equality to all $H(\kappa)$ it suffices to observe
that, by the Krein formula, the resolvent of $H(\kappa)$ is a rank-one
perturbation of the resolvent of $H(\infty)$. \end{proof}

By the general theory of Sturm-Liouville operators, the resolvent
kernels $\mathcal{G}_{m}(z;\varrho,\varrho')$ of the partial Hamiltonians
$H_{0}(\infty)$, if $m=0$, and $H_{m}=H_{\mathrm{min},m}=H_{\mathrm{max},m}$,
if $m\neq0$, are equal to \begin{align*}
\mathcal{G}_{m}(z;\varrho,\varrho') & =\frac{1}{2(2|m|)!\,\sqrt{-z}\,\sqrt{\varrho\varrho'}}\,\Gamma\!\left(\frac{1}{2}+|m|-\frac{Z}{2\sqrt{-z}}\right)\\
\noalign{\medskip} & \hspace{1em}\times\, M_{Z/(2\sqrt{-z}),|m|}\!\left(2\sqrt{-z}\varrho_{<}\right)W_{Z/(2\sqrt{-z}),|m|}\!\left(2\sqrt{-z}\varrho_{>}\right)\end{align*}
Here $\varrho_{<}$, $\varrho_{>}$ denote the smaller and the greater
out of $\varrho$, $\varrho'$, respectively.

The Green function $\mathcal{G}_{0}^{\kappa}(z;\varrho,\varrho')$
for the Hamiltonian $H_{0}(\kappa)$, $\kappa\in\mathbb{R}$, can
be constructed using the Krein resolvent formula that guarantees existence
of a function $\phi^{\kappa}(z)$ such that\begin{equation}
\mathcal{G}_{0}^{\kappa}(z;\varrho,\varrho')=\mathcal{G}_{0}(z;\varrho,\varrho')+\frac{\phi^{\kappa}(z)}{\sqrt{\varrho\varrho'}}\, W_{Z/(2\sqrt{-z}),0}\!\left(2\sqrt{-z}\varrho\right)W_{Z/(2\sqrt{-z}),0}\!\left(2\sqrt{-z}\varrho'\right),\label{eq:Green0}\end{equation}
with $z\in\mathbb{C}\setminus\mathbb{R}$. Since the integral kernel
must satisfy the same boundary condition as that defining $\Dom H_{0}(\kappa)$
we have \begin{equation}
\phi^{\kappa}(z)=\frac{1}{2\sqrt{-z}}\,\Gamma\!\left(\frac{1}{2}-\frac{Z}{2\sqrt{-z}}\right)^{\!2}\left(2\gamma+\ln(2\sqrt{-z})+\Psi\!\left(\frac{1}{2}-\frac{Z}{2\sqrt{-z}}\right)+\kappa\right)^{\!-1}\label{eq:phi_kappa}\end{equation}
where $\Psi(z)=\Gamma'(z)/\Gamma(z)$ is the polygamma function and
$\gamma=-\Gamma'(1)$ is the Euler constant. The Green function of
$H(\kappa)$, $\kappa\in\mathbb{R}\cup\{\infty\}$, expressed in polar
coordinates, equals \begin{equation}
\begin{split}\mathcal{G}^{\kappa}(z;\varrho,\varphi,\varrho',\varphi')=\, & \frac{1}{2\pi}\sum_{m=-\infty}^{\infty}\mathcal{G}_{m}(z;\varrho,\varrho')\, e^{im(\varphi-\varphi')}\\
\noalign{\smallskip} & +\frac{\phi^{\kappa}(z)}{2\pi\sqrt{\varrho\varrho'}}\, W_{Z/(2\sqrt{-z}),0}\!\left(2\sqrt{-z}\varrho\right)W_{Z/(2\sqrt{-z}),0}\!\left(2\sqrt{-z}\varrho'\right).\end{split}
\label{eq:green}\end{equation}

The point spectrum of $H_{C}$ equals the union of the point spectra
of $H_{m}$, $m\in\mathbb{Z}$ (with $H_{0}\equiv H_{0}(\infty)$).
The eigenvalues of $H_{C}$ jointly with eigenfunctions are computed
in \cite{ygc} and correspond to the poles of the respective Green
functions. Thus we recall that all eigenvalues of $H_{m}$, $m\in\mathbb{Z}$,
are simple and are equal to \begin{equation}
\lambda_{m,n}=-\frac{Z^{2}}{(2|m|+2n+1)^{2}}\,,\ n\in\mathbb{Z}_{+},\label{eq:ev}\end{equation}
(here $\mathbb{Z}_{+}=\{n\in\mathbb{Z};n\geq0\}$). Denote by $N=|m|+n+1$
the principal quantum number and put $\lambda_{N}=\lambda_{m,n}$
for $|m|+n=N-1$, i.e. $\lambda_{N}=-Z^{2}/(2N-1)^{2}$, $N\in\mathbb{N}$.
Then the multiplicity of $\lambda_{N}$ in the spectrum of $H_{C}$
is $2N-1$. The corresponding normalized eigenfunctions are \begin{align*}
\psi_{m,n}(\varrho,\varphi)= & \left(\frac{n!}{2\pi\,(n+2|m|)!}\right)^{\!1/2}\frac{2Z}{(2|m|+2n+1)^{3/2}}\left(\frac{2Z\varrho}{2|m|+2n+1}\right)^{\!|m|}\\
\noalign{\smallskip} & \times\, L_{n}^{(2|m|)}\!\left(\frac{2Z\varrho}{2|m|+2n+1}\right)e^{-Z\varrho/(2|m|+2n+1)}\, e^{im\varphi},\end{align*}
 where $L_{n}^{(2|m|)}$ stands for the associated Laguerre polynomial.

Using a similar reasoning as in \cite{bgl} one concludes that the
point spectrum of $H(\kappa)$, $\kappa\in\mathbb{R}$, contains the
eigenvalues $\lambda_{N}$ with multiplicities $2(N-1)$ (hence $\lambda_{1}$
is missing). In fact, the point spectrum of $H_{m}$ is simple and
is formed by the eigenvalues $\lambda_{N}$, $N\geq|m|+1$. If a point
interaction is switched on then the spectrum of the component $H_{0}$
is deformed while the point spectra of the components $H_{m}$, $m\neq0$,
remain untouched. On the other hand, if $\kappa\in\mathbb{R}$ then
additional eigenvalues emerge in the spectrum of $H(\kappa)$, the
so called point levels. They are simple and negative. Let us denote
them in ascending order by $\epsilon_{j}(Z;\kappa)$, $j=\mathbb{Z}_{+}$.

From the general theory concerned with Friedrichs extensions \cite[Theorem~X.23]{rs2}
and location of discrete spectra of self-adjoint extensions \cite[Chp.~8.3]{weidmann}
one deduces that the points levels are located as follows\[
\epsilon_{0}(Z;\kappa)<\lambda_{1}<\epsilon_{1}(Z;\kappa)<\lambda_{2}<\epsilon_{2}(Z;\kappa)<\lambda_{3}<\ldots<0.\]
Using the substitution\[
\epsilon_{j}(Z;\kappa)=-Z^{2}k_{j}(\kappa)^{2},\ \kappa_{0}=\kappa+\ln Z,\]
one finds from (\ref{eq:Green0}) and (\ref{eq:phi_kappa}) that the
equation on point levels takes the form \begin{equation}
2\gamma+\ln(2k)+\Psi\!\left(\frac{1}{2}-\frac{1}{2k}\right)+\kappa_{0}=0,\label{eq:eq_point_levs}\end{equation}
with the unknown $k=k_{j}(\kappa)>0$. This implies the scaling property
\[
\epsilon_{j}(Z;\kappa)=Z^{2}\,\epsilon_{j}(1;\kappa+\ln Z),\ j=0,1,2,\ldots.\]
By an elementary analysis of equation (\ref{eq:eq_point_levs}) one
can show that $\epsilon_{j}(Z;\kappa)$ are strictly increasing functions
in the parameter $\kappa\in\mathbb{R}$, and one has the asymptotic
formulas\[
\epsilon_{j}(Z;\kappa)=-\frac{Z^{2}}{(2j+1)^{2}}-\frac{4Z^{2}}{(2j+1)^{3}\kappa}+O(\kappa^{-2})\quad\mbox{as }\kappa\to+\infty,\]
for all $j\geq0$, and\begin{eqnarray*}
\epsilon_{0}(Z;\kappa) & = & -4\, e^{-2\gamma-2\kappa}+O(e^{-\kappa})\quad\mbox{as }\kappa\to-\infty,\\
\epsilon_{j}(Z;\kappa) & = & -\frac{Z^{2}}{(2j-1)^{2}}-\frac{4Z^{2}}{(2j-1)^{3}\kappa}+O(\kappa^{-2})\quad\mbox{as }\kappa\to-\infty,\mbox{ }j\geq1.\end{eqnarray*}
Figure~1 depicts several first point levels as functions of $\kappa$
for $Z=1$.

Finally note that from the form of the Green function one can also
derive normalized eigenfunctions corresponding to the point levels,
namely \begin{align*}
\eta_{j}(\kappa;\varrho,\varphi)=\, & \sqrt{\frac{Z}{2\pi\varrho}}\, k_{j}(\kappa)\left(k_{j}(\kappa)+\frac{1}{2}\,\Psi'\!\left(\frac{1}{2}-\frac{1}{2k_{j}(\kappa)}\right)\right)^{\!-1/2}\Gamma\!\left(\frac{1}{2}-\frac{1}{2k_{j}(\kappa)}\right)\\
\noalign{\smallskip} & \times\, W_{1/(2k_{j}(\kappa)),0}\big(2k_{j}(\kappa)Z\varrho\big),\end{align*}
where $k_{j}(\kappa)=(-\epsilon_{j}(Z;\kappa))^{1/2}/Z$, $j=0,1,2,\ldots$.

\subsection{The momentum representation}

The normalized generalized eigenfunctions for $H_{m}$, $m\in\mathbb{Z}$,
(with $H_{0}\equiv H_{0}(\infty)$) are known including the correct
normalization \cite{ygc}. One has, with $k>0$,\[
\psi_{m}(k,\varrho)=\frac{1}{(2|m|)!}\!\left(\!\frac{2}{1+e^{-\pi Z/k}}\!\right)^{\!\!1/2}\,\prod_{s=0}^{|m|-1}\!\left(\!\left(\! s+\frac{1}{2}\right)^{\!2}+\frac{Z^{2}}{4k^{2}}\right)^{\!\!1/2}\frac{i^{m}}{\sqrt{2ik\varrho}}\, M_{Z/(2ik),|m|}(2ik\varrho).\]
In a comparatively recent paper \cite{GesztesyZinchenko} a large
class of Schr\"odinger operators on a halfline with strongly singular
potentials is studied, with the results being directly applicable
to operators $H_{m}$, $m\in\mathbb{Z}$. In that article, a measure
on the dual space is constructed with the aid of the associated Titchmarsh-Weyl
$m$-function, and unitarity of the eigenfunction expansion, involving
both proper and generalized eigenfunctions, is proven (one can also
consult paper \cite{Fulton} which appeared later on and covers a
less general class of potentials but with our example still being
included). As a consequence one deduces that the integral transform\[
\sT_{m}:L^{2}(\mathbb{R}_{+},\varrho\,\mbox{d}\varrho)\to L^{2}(\mathbb{R}_{+},k\,\mbox{d}k),\mbox{ }\sT_{m}[f](k)=\int_{0}^{\infty}\psi_{m}(k,\varrho)f(\varrho)\varrho\,\mbox{d}\varrho\]
is a well defined bounded operator. Denote by $\sH_{m,\mathrm{pp}}$
the closure of the subspace in $L^{2}(\mathbb{R}_{+},\varrho\,\mbox{d}\varrho)$
spanned by the eigenfunctions $\psi_{m,n}(\varrho)$, $n\in\mathbb{Z}_{+}$,
and by $\sH_{m,\mathrm{ac}}$ its orthogonal complement. Then the
kernel of $\sT_{m}$ equals $\sH_{m,\mathrm{pp}}$, and the restriction\[
\sT_{m,\mathrm{ac}}:=\sT_{m}\big|_{\sH_{m,\mathrm{ac}}}:\sH_{m,\mathrm{ac}}\to L^{2}(\mathbb{R}_{+},k\,\mbox{d}k)\]
is a unitary mapping. Moreover, $\sT_{m,\mathrm{ac}}$ transforms
$H_{m}\big|_{\sH_{m,\mathrm{ac}}}$ into the multiplication operator
by the function $k^{2}$ acting in $L^{2}(\mathbb{R}_{+},k\,\mbox{d}k)$.
It follows that the essential spectrum of $H_{m}$ is in fact absolutely
continuous. The same is also true for all $H_{0}(\kappa)$, $\kappa\in\mathbb{R}$.

\begin{proposition} For all $m\in\mathbb{Z}$ one has $\sigma_{\mathrm{ess}}(H_{m})=\sigma_{\mathrm{ac}}(H_{m})=[\,0,\infty)$
and $\sigma_{\mathrm{pp}}(H_{m})=\overline{\{\lambda_{m,n};\, n\in\mathbb{Z}_{+}\}}$
(with the only accumulation point being just $0$). Similarly, for
all $\kappa\in\mathbb{R}$ one has $\sigma_{\mathrm{ess}}(H_{0}(\kappa))=\sigma_{\mathrm{ac}}(H_{0}(\kappa))=[\,0,\infty)$
and $\sigma_{\mathrm{pp}}(H_{0}(\kappa))=\overline{\{\epsilon_{j}(\kappa);\, j\in\mathbb{Z}_{+}\}}$.
Moreover, the spectra of $H_{m}$, $m\in\mathbb{Z}$, and $H_{0}(\kappa)$,
$\kappa\in\mathbb{R}$, are simple. In particular, the singular continuous
spectra of $H_{m}$ and $H_{0}(\kappa)$ are empty. \end{proposition}

The transformation inverse to $\sT_{m,\mathrm{ac}}$ is\[
\sT_{m,\mathrm{ac}}^{\,-1}:L^{2}(\mathbb{R}_{+},k\,\mbox{d}k)\to\sH_{m,\mathrm{ac}},\ \sT_{m,\mathrm{ac}}^{\,-1}[g](\varrho)=\int_{0}^{\infty}\psi_{m}(k,\varrho)g(k)k\,\mbox{d}k.\]
Thus one concludes that $H_{m}$ in $L^{2}(\mathbb{R}_{+},\varrho\,\mbox{d}\varrho)$
is unitarily equivalent to $\hat{H}_{m}$ in\[
\hat{\sH}_{m}=\ell^{2}(\mathbb{Z}_{+})\oplus L^{2}(\mathbb{R}_{+},k\,\mbox{d}k).\]
$\Dom(\hat{H}_{m})$ is formed by those $\hat{f}=\{\hat{f}_{n}\}_{n=0}^{\infty}+\hat{f}(k)\in\hat{\sH}_{m}$
for which $k^{2}\hat{f}(k)\in L^{2}(\mathbb{R}_{+},k\,\mbox{d}k)$.
If $\hat{f}\in\Dom(\hat{H}_{m})$ then\[
\hat{H}_{m}\hat{f}=\{\lambda_{m,n}\hat{f}_{n}\}_{n=0}^{\infty}+k^{2}\hat{f}(k).\]
The unitary mapping $\hat{\sH}_{m}\to L^{2}(\mathbb{R}_{+},\varrho\,\mbox{d}\varrho):\hat{f}\mapsto f$
is given by\[
f(\varrho)=\sum_{n=0}^{\infty}\hat{f}_{n}\psi_{m,n}(\varrho)+\int_{0}^{\infty}\psi_{m}(k,\varrho)\hat{f}(k)k\,\mbox{d}k.\]
Conversely, $\hat{f}_{n}=\langle\psi_{m,n},f\rangle$, $\mbox{ }\hat{f}(k)=\sT_{m}[f](k)$.
One has\[
\|\hat{f}\|^{2}=\sum_{n=0}^{\infty}|\hat{f}_{n}|^{2}+\int_{0}^{\infty}|\hat{f}(k)|^{2}k\,\mbox{d}k=\int_{0}^{\infty}|f(\varrho)|^{2}\varrho\,\mbox{d}\varrho=\|f\|^{2}.\]

One can use the momentum representation for an alternative and equivalent
construction of point interactions. It again turns out that an nontrivial
result can be derived only in the sector $m=0$ to which we confine
our attention. A symmetric restriction $A$ of $\hat{H}_{0}$ is obtained
by requiring that $f(0)=0$ if $\hat{f}\in\Dom A\subset\Dom\hat{H}_{0}$.
More details follow. From now on we omit, in the notation, the hat
over elements $f\in\hat{\sH}_{0}$.

Let us denote the normalization factor of generalized eigenfunctions
as\[
N(k)=\left(\frac{2}{1+e^{-\pi Z/k}}\right)^{\!1/2},\mbox{ }k>0.\]
For $g\in\hat{\sH}_{0}$ such that $g(k)\in L^{1}(\mathbb{R}_{+},k\mbox{d}k)$
put\[
S(g)=\sum_{n=0}^{\infty}\frac{2Z}{(2n+1)^{3/2}}\, g_{n}+\int_{0}^{\infty}N(k)g(k)k\,\mbox{d}k.\]
For $\xi\in\mathbb{C}$ and $f\in\hat{\sH}_{0}$ such that $f(k)-\xi N(k)/(k^{2}+Z^{2})\in L^{1}(\mathbb{R}_{+},k\mbox{d}k)$
put\[
S(\xi,f)=\sum_{n=0}^{\infty}\frac{2Z}{(2n+1)^{3/2}}\, f_{n}+\int_{0}^{\infty}N(k)\left(f(k)-\frac{\xi N(k)}{k^{2}+Z^{2}}\right)k\,\mbox{d}k.\]
Clearly, if $\xi$ exists then it is unambiguously determined by $f$,
and $S(g)\equiv S(0,g)$. Observe that $\forall g\in\Dom(\hat{H}_{0})$,
$g(k)\in L^{1}(\mathbb{R}_{+},k\mbox{d}k)$, and one has $\check{g}(0)=S(g)$
where\begin{equation}
\check{g}(\varrho)=\sum_{n=0}^{\infty}g_{n}\psi_{0,n}(\varrho)+\int_{0}^{\infty}\psi_{0}(k,\varrho)g(k)k\,\mbox{d}k.\label{eq:g_hacek}\end{equation}

One defines $A\subset\hat{H}_{0}$ by\[
\Dom(A)=\{g\in\Dom(\hat{H}_{0});\mbox{ }S(g)=0\}.\]
It is not difficult to check that $f\in\Dom(A^{\ast})$ if and only
if there exist (necessarily unique) $\xi\in\mathbb{C}$ and $\eta\in\hat{\sH}_{0}$
such that for all $n\in\mathbb{Z}_{+}$ and almost all $k>0$,\[
\lambda_{0,n}f_{n}=\eta_{n}+\frac{2Z\xi}{(2n+1)^{3/2}}\,,\mbox{ }k^{2}f(k)=\eta(k)+\xi N(k).\]
In that case, $A^{\ast}f=\eta$. Note that if $f\in\Dom(A^{\ast})$
and $\xi$, $\eta$ are as above then\[
f(k)-\frac{\xi N(k)}{k^{2}+Z^{2}}=\frac{Z^{2}f(k)}{k^{2}+Z^{2}}+\frac{\eta(k)}{k^{2}+Z^{2}}\in L^{1}(\mathbb{R}_{+},k\mbox{d}k),\]

Let us now discuss self-adjoint extensions of $A$. The deficiency
indices of $A$ are $(1,1)$. For $z\in\mathbb{C}\setminus\mathbb{R}$,
$\Ker(A^{\ast}-z)=\mathbb{C}f_{z}$ where\begin{equation}
\forall n\in\mathbb{Z}_{+},\mbox{ }(f_{z})_{n}=\frac{2Z}{(2n+1)^{3/2}(\lambda_{0,n}-z)}\,;\mbox{ }\forall k>0,\mbox{ }\mbox{ }f_{z}(k)=\frac{N(k)}{k^{2}-z}\,.\label{eq:fz_basis_fce_defic}\end{equation}
For the computational convenience the spectral parameter is chosen
to be $z=iZ^{2}/2$. For $e^{i\alpha}\in\mathbb{T}^{1}$ let $A_{\alpha}$
be the self-adjoint extension of $A$ defined by\begin{eqnarray*}
 &  & \Dom(A_{\alpha})\,=\,\Dom(A)+\mathbb{C}\left(f_{z}+e^{i\alpha}f_{\bar{z}}\right),\\
 &  & A_{\alpha}\!\left(g+t\left(f_{z}+e^{i\alpha}f_{\bar{z}}\right)\right)=\, Ag+t\left(zf_{z}+\bar{z}e^{i\alpha}f_{\bar{z}}\right).\end{eqnarray*}
If\begin{equation}
f=g+t\left(f_{z}+e^{i\alpha}f_{\bar{z}}\right)\in\Dom(A_{\alpha}),\mbox{ with }g\in\Dom(A),\mbox{ }t\in\mathbb{C},\label{eq:f_DomAalpha}\end{equation}
then there exists $\xi\in\mathbb{C}$ (necessarily unique) such that\[
f(k)-\frac{\xi N(k)}{k^{2}+Z^{2}}\in L^{1}(\mathbb{R}_{+},k\mbox{d}k),\]
namely $\xi=t\,(1+e^{i\alpha})$. Furthermore, one has\[
S(\xi,f)=t\left(S(1,f_{z})+e^{i\alpha}S(1,f_{\bar{z}})\right).\]
A straightforward computation gives\[
S(1,f_{z})=-\Re\!\left(\Psi\!\left(\frac{i}{2}\right)\right)-\gamma-\frac{7}{2}\ln(2)+i+\frac{i\pi}{4}+\frac{i\pi}{2}\coth\!\left(\frac{\pi}{2}\right).\]
The computation is based on the following identities: for $a\notin2\mathbb{Z}_{+}+1$,\begin{equation}
\sum_{n=0}^{\infty}\frac{1}{(2n+1)(2n+1-a)}=\frac{1}{2a}\left(\Psi\!\left(\frac{1}{2}\right)-\Psi\!\left(\frac{1-a}{2}\right)\right),\label{eq:sum_nna}\end{equation}
and, for $a\notin(-\infty,0\,]$,\begin{equation}
\int_{0}^{\infty}\frac{y}{(1+e^{\pi y})(y^{2}+a)}\,\mbox{d}y=-\frac{1}{4}\ln(4a)+\Psi(\sqrt{a}\,)-\frac{1}{2}\,\Psi\!\left(\frac{\sqrt{a}}{2}\right).\label{eq:int_y2a}\end{equation}
Notice that $S(1,f_{\bar{z}})=\overline{S(1,f_{z})}$. Put\begin{eqnarray*}
\hat{\kappa} & = & \frac{1}{1+e^{i\alpha}}\left(S(1,f_{z})+e^{i\alpha}\overline{S(1,f_{z})}\right)\\
 & = & -\Re\!\left(\Psi\!\left(\frac{i}{2}\right)\right)-\gamma-\frac{7}{2}\,\ln(2)+\left(1+\frac{\pi}{4}+\frac{\pi}{2}\coth\!\left(\frac{\pi}{2}\right)\right)\tan\!\left(\frac{\alpha}{2}\right).\end{eqnarray*}
Still assuming (\ref{eq:f_DomAalpha}) one has $S(\xi,f)=\xi\hat{\kappa}.$
Let us redenote $A_{\alpha}=\hat{H}_{0}(\hat{\kappa})$.

One concludes that the one-parameter family of all self-adjoint extensions
of $A$ is $\hat{H}_{0}(\hat{\kappa})$, $\hat{\kappa}\in\mathbb{R}\cup\{\infty\}$.
A vector $f\in\hat{\sH}_{0}$ belongs to $\Dom(\hat{H}_{0}(\hat{\kappa}))$
iff there exists $\xi\in\mathbb{C}$ (necessarily unique) such that\begin{equation}
f(k)-\frac{\xi N(k)}{k^{2}+Z^{2}}\in L^{1}(\mathbb{R}_{+},k\mbox{d}k),\ k^{2}f(k)-\frac{1}{\hat{\kappa}}\, S(\xi,f)N(k)\in L^{2}(\mathbb{R}_{+},k\mbox{d}k).\label{eq:C2_def_DomHg}\end{equation}
Then\[
\hat{H}_{0}(\hat{\kappa})f=\left\{ \lambda_{0,n}f_{n}-\frac{2Z\xi}{(2n+1)^{3/2}}\right\} _{n=0}^{\infty}+\left(k^{2}f(k)-\frac{1}{\hat{\kappa}}\, S(\xi,f)N(k)\right).\]
In addition one has $S(\xi,f)/\hat{\kappa}=\xi$. Clearly, $\hat{H}_{0}(\infty)=\hat{H}_{0}$.

Let us check the point spectrum of $\hat{H}_{0}(\hat{\kappa})$, $\hat{\kappa}\in\mathbb{R}$.
Suppose $0\neq f\in\hat{\sH}_{0}$ and $\lambda\in\mathbb{R}$ fulfill
$\hat{H}_{0}(\hat{\kappa})f=\lambda f$. This means that\begin{equation}
\lambda_{0,n}f_{n}-\frac{2Z\xi}{(2n+1)^{3/2}}=\lambda f_{n}\mbox{ for }n\in\mathbb{Z}_{+},\mbox{ }k^{2}f(k)-\xi N(k)=\lambda f(k)\mbox{ for }k>0.\label{eq:eigenval_Cg_D2}\end{equation}
Clearly, $\lambda$ must be negative since otherwise $f(k)=N(k)/(k^{2}-\lambda)$
would not be $L^{2}$ integrable. Furthermore, the point spectrum
of $\hat{H}_{0}(\hat{\kappa})$ is disjoint with the point spectrum
of $\hat{H}_{0}$. In fact, suppose $\lambda=\lambda_{0,p}$ for some
$p\in\mathbb{Z}_{+}$. Then from the first equation in (\ref{eq:eigenval_Cg_D2}),
with $n=p$, it follows that $\xi=0$. Moreover, (\ref{eq:eigenval_Cg_D2})
implies $f_{n}=0$ for $n\neq p$, and $f(k)=0$ for $k>0$. Necessarily,
$f_{p}\neq0$. Then \[
S(\xi,f)=2Z(2p+1)^{-3/2}f_{p}\neq0\]
 and the second condition in (\ref{eq:C2_def_DomHg}) is not satisfied,
a contradiction.

Suppose $\lambda<0$ and $\lambda\neq\lambda_{0,n}$, $\forall n$,
is an eigenvalue. Then there exists one independent eigenvector $f$
corresponding to $\lambda$ for which one can put $\xi=1$,\begin{equation}
f_{n}=\frac{2Z}{(2n+1)^{3/2}(\lambda_{0,n}-\lambda)}\quad\mbox{for }n\in\mathbb{Z}_{+},\mbox{ }f(k)=\frac{N(k)}{k^{2}-\lambda}\quad\mbox{for }k>0.\label{eq:C2_eigenv_f}\end{equation}
The eigenvalue equation reads $S(1,f)=\hat{\kappa}$, with $f$ given
in (\ref{eq:C2_eigenv_f}), i.e.\[
\sum_{n=0}^{\infty}\frac{4Z^{2}}{(2n+1)^{3}(\lambda_{0,n}-\lambda)}+\int_{0}^{\infty}\frac{2}{1+e^{-\pi Z/k}}\left(\frac{1}{k^{2}-\lambda}-\frac{1}{k^{2}+Z^{2}}\right)k\,\mbox{d}k=\hat{\kappa}.\]
One can get rid of the parameter $Z$ using the substitution $\lambda=-Z^{2}/x^{2}$.
With the aid of (\ref{eq:sum_nna}) and (\ref{eq:int_y2a}) one finds
that $\lambda=-Z^{2}/x^{2}$ is an eigenvalue iff $x$ solves the
equation\begin{equation}
\pi\tan\!\left(\frac{\pi}{2}\, x\right)+\ln(x)-\Psi\!\!\left(\frac{1+x}{2}\right)-\gamma-4\ln(2)=\hat{\kappa}.\label{eq:eigenvaleq_x}\end{equation}

\subsection{A relation between the two representations}

We wish to compare the operators $\hat{H}_{0}(\hat{\kappa})$ and
$H_{0}(\kappa)$. The domain of the latter Hamiltonian in the coordinate
representation is given by a boundary condition at the origin. So
we have to determine the asymptotic behavior of $\check{g}(\varrho)$
as $\varrho\to0$ for an arbitrary $g\in\Dom\hat{H}_{0}(\hat{\kappa})$,
with $\check{g}$ being given in (\ref{eq:g_hacek}).

As a first step we find a relation between the basis function $f_{z}$
of the deficiency subspace given in (\ref{eq:fz_basis_fce_defic})
and $\check{f}_{z}(\varrho)$, a basis function of the deficiency
subspace in the coordinate representation. To simplify the notation
let us temporarily set $Z=1$. We put\[
\check{f}_{z}(\varrho)=\frac{1}{\sqrt{\varrho}}\, W_{1/(2\sqrt{-z}),0}(2\sqrt{-z}\,\varrho),\mbox{ }z\in\mathbb{C}\setminus[\,0,+\infty).\]
This can be rewritten in terms of confluent hypergeometric functions,\[
\check{f}_{z}(\varrho)=\sqrt{2}\,(-z)^{1/4}\, e^{-\sqrt{-z}\,\varrho}\, U\!\!\left(\frac{1}{2}-\frac{1}{2\sqrt{-z}},1,2\sqrt{-z}\,\varrho\right).\]
One knows that\[
\sum_{n=0}^{\infty}(f_{z})_{n}\psi_{0,n}(\varrho)+\int_{0}^{\infty}f_{z}(k)\psi_{0}(k,\varrho)k\,\mbox{d}k=C(z)\check{f}_{z}(\varrho)\]
where $C(z)$ is a holomorphic function on $\mathbb{C}\setminus[\,-1,+\infty)$.

By unitarity,\[
\sum_{n=0}^{\infty}|(f_{z})_{n}|^{2}+\int_{0}^{\infty}|f_{z}(k)|^{2}k\,\mbox{d}k=|C(z)|^{2}\int_{0}^{\infty}|\check{f}_{z}(\varrho)|^{2}\varrho\,\mbox{d}\varrho.\]
Suppose $z<-1$. In that case,\[
\int_{0}^{\infty}\check{f}_{z}(\varrho)^{2}\varrho\mbox{d}\varrho=\frac{2\sqrt{-z}+\Psi'\!\!\left(\frac{1}{2}-\frac{1}{2\sqrt{-z}}\right)}{2(-z)\Gamma\!\!\left(\frac{1}{2}-\frac{1}{2\sqrt{-z}}\right)^{\!2}}.\]
Furthermore,\[
\sum_{n=0}^{\infty}(f_{z})_{n}^{\,\,2}=\frac{1}{4(-z)^{3/2}}\!\left(\Psi'\!\!\left(\frac{1}{2}-\frac{1}{2\sqrt{-z}}\right)-\Psi'\!\!\left(\frac{1}{2}+\frac{1}{2\sqrt{-z}}\right)\!\right)\!.\]
Using the identity\[
\int_{0}^{\infty}\frac{1}{\cosh(\pi x/2)^{2}(x^{2}+a^{2})}\,\mbox{d}x=\frac{1}{\pi a}\,\Psi'\!\!\left(\frac{1+a}{2}\right)\!\]
one finds that\[
\int_{0}^{\infty}f_{z}(k)^{2}k\,\mbox{d}k=-\frac{1}{2z}+\frac{1}{4(-z)^{3/2}}\,\Psi'\!\!\left(\frac{1}{2}+\frac{1}{2\sqrt{-z}}\right)\!.\]
Finally one arrives at the equality\begin{align}
 & \sum_{n=0}^{\infty}(f_{z})_{n}\psi_{0,n}(\varrho)+\int_{0}^{\infty}f_{z}(k)\psi_{0}(k,\varrho)k\,\mbox{d}k\nonumber \\
 & =\frac{1}{(-z)^{1/4}}\,\Gamma\!\!\left(\frac{1}{2}-\frac{1}{2\sqrt{-z}}\right)\frac{1}{\sqrt{2\varrho}}\, W_{\frac{1}{2\sqrt{-z}},0}(2\sqrt{-z}\,\varrho).\label{eq:transf_fz_fzhacek}\end{align}

Using a simple scaling one can return back to a general parameter
$Z>0$. Considering the limit $z\to-1$ in (\ref{eq:transf_fz_fzhacek})
one derives the asymptotic formula\[
\int_{0}^{\infty}\psi_{0}(k,\varrho)\,\frac{N(k)}{k^{2}+Z^{2}}\, k\,\mbox{d}k=-\ln(Z\varrho)-\gamma+3\ln(2)+O(\varrho\ln(\varrho))\quad\mbox{as }\varrho\to0.\]
Suppose $f\in\Dom(\hat{H}_{0}(\hat{\kappa}))$. Then ($\xi\in\mathbb{C}$
is introduced in the definition of $\Dom(\hat{H}_{0}(\hat{\kappa}))$)\[
\check{f}(\varrho)=\sum_{n=0}^{\infty}f_{n}\psi_{0,n}(\varrho)+\int_{0}^{\infty}\psi_{0}(k,\varrho)\!\left(\! f(k)-\frac{\xi N(k)}{k^{2}+Z^{2}}\right)\! k\mbox{d}k+\xi\int_{0}^{\infty}\psi_{0}(k,\varrho)\,\frac{N(k)}{k^{2}+Z^{2}}\, k\,\mbox{d}k.\]
Hence\[
\check{f}(\varrho)=S(\xi,f)+\xi(-\ln(Z\varrho)-\gamma+3\ln(2))+o(1)\quad\mbox{as }\varrho\to0.\]
Recall that $S(\xi,f)=\xi\hat{\kappa}$. One concludes that\begin{equation}
\forall f\in\Dom(\hat{H}_{0}(\hat{\kappa})),\mbox{ }\check{f}(\varrho)=\xi\big(-\ln(Z\varrho)+\hat{\kappa}-\gamma+3\ln(2)\big)+o(1)\quad\mbox{as }\varrho\to0.\label{eq:asympt_r0_fhacek}\end{equation}

Since the domain of $H_{0}(\kappa)$ is determined by the asymptotic
behavior at $\varrho=0$,\begin{equation}
\check{f}(\varrho)=-\alpha_{0}\ln(\varrho)+\alpha_{1}+o(1)\mbox{ }\mbox{ as }\varrho\to0,\mbox{ where }\alpha_{1}=\kappa\alpha_{0},\label{eq:asympt_r0_PhD}\end{equation}
one finds, by comparing (\ref{eq:asympt_r0_fhacek}) and (\ref{eq:asympt_r0_PhD}),
that the operators $H_{0}(\kappa)$ and $\hat{H}_{0}(\hat{\kappa})$
are unitarily equivalent if\[
\kappa=\hat{\kappa}-\ln(Z)-\gamma+3\ln(2).\]

\section{A Hydrogen atom in a thin layer \label{sec:layer}}

\subsection{Notation}

We wish to discuss a model describing a Hydrogen atom or a Hydrogen-like
ion confined to an infinite planar slab $\Omega_{a}$ of width $a$.
Thus we denote \[
\Omega_{a}=\mathbb{R}^{2}\times\left(-\frac{a}{2}\,,\,\frac{a}{2}\right)\subset\mathbb{R}^{3}.\]
Our goal is to consider the limit when the width $a$ tends to zero.
Let us first introduce the notation and recall some related results.

For $\Omega\subset\mathbb{R}^{n}$, an nonempty open set with a Lipschitz
continuous boundary on each component, denote by $\mathcal{H}^{m}(\Omega)$
the $m$th Sobolev space and by $\mathcal{H}_{0}^{m}(\Omega)$ the
closure of $C_{0}^{\infty}(\Omega)$ in $\mathcal{H}^{m}(\Omega)$.
One has a natural isometric embedding $\mathcal{H}_{0}^{m}(\Omega)\subset\mathcal{H}^{m}(\mathbb{R}^{n})$.
Furthermore, $\mathcal{D}^{1,2}(\mathbb{R}^{n})$ denotes the completion
of $C_{0}^{\infty}(\mathbb{R}^{n})$ with respect to the norm \[
u\mapsto\left(\int_{\mathbb{R}^{n}}|\nabla u|^{2}\,\mathrm{d}\mathbf{x}\right)^{\!1/2}.\]
In this case one has a continuous embedding $\mathcal{H}^{1}(\mathbb{R}^{n})\subset\mathcal{D}^{1,2}(\mathbb{R}^{n})$.
Recall also that the Dirichlet Laplacian $-\Delta_{D}$ is the unique
self-adjoint operator associated with the closed positive form $q(f,g)=\langle\nabla f,\nabla g\rangle$
defined on $\mathcal{H}_{0}^{1}(\Omega)$ (the scalar product is taken
in $L^{2}(\Omega,\mbox{d}\mathbf{x})$). The form representation theorem
implies that $\Dom\Delta_{D}=\mathcal{H}_{0}^{1}(\Omega)\cap\mathcal{H}^{2}(\Omega)$.

Below we employ the Hardy inequality in $\mathbb{R}^{3}$ which states
that for any $u\in\mathcal{D}^{1,2}(\mathbb{R}^{3})$, \begin{equation}
\frac{1}{4}\int_{\mathbb{R}^{3}}\frac{|u(\mathbf{x})|^{2}}{|\mathbf{x}|^{2}}\,\mathrm{d}\mathbf{x}\leq\int_{\mathbb{R}^{3}}|\nabla u(\mathbf{x})|^{2}\,\mathrm{d}\mathbf{x}.\label{eq:Hardy}\end{equation}
The Hardy inequality is extended to domains with boundaries in \cite{WangZhu}
where one can find additional references. In the case of the Dirichlet
boundary condition, however, one can simply make use of the chain
of embeddings $\mathcal{H}_{0}^{1}(\Omega)\subset\mathcal{H}^{1}(\mathbb{R}^{3})\subset\mathcal{D}^{1,2}(\mathbb{R}^{3})$.
Hence inequality (\ref{eq:Hardy}) holds for any $u\in\mathcal{H}_{0}^{1}(\Omega)$
where $\Omega\subset\mathbb{R}^{3}$ is still supposed to have the
above stated properties.

In our model we introduce the Hamiltonian \begin{equation}
H^{a}=-\Delta_{D}-\frac{Z}{r}\,,\label{eq:Ha_def}\end{equation}
with $r=\sqrt{x^{2}+y^{2}+z^{2}}$ and $Z>0$, in the Hilbert space
$L^{2}(\Omega_{a})$. To see that a self-adjoint operator is well
defined by relation (\ref{eq:Ha_def}) it suffices to show that the
potential $1/r$ is $(-\Delta_{D})$ bounded with a relative bound
less than one (or even zero) and to refer to the Kato-Rellich theorem.
In fact, by the Hardy inequality (\ref{eq:Hardy}), the estimate \[
\|r^{-1}\psi\|^{2}\leq4\|\nabla\psi\|^{2}=4\langle\psi,-\Delta_{D}\psi\rangle\leq2\!\left(\epsilon^{2}\|-\Delta_{D}\psi\|^{2}+\frac{1}{\epsilon^{2}}\,\|\psi\|^{2}\right)\]
holds for all $\psi\in\Dom\Delta_{D}$ and $\epsilon>0$. Thus one
has\[
\Dom H^{a}=\Dom(-\Delta_{D})=\mathcal{H}_{0}^{1}(\Omega_{a})\cap\mathcal{H}^{2}(\Omega_{a}),\ \, Q(H^{a})=\mathcal{H}_{0}^{1}(\Omega_{a})\]
(here $Q(A)$ stands for the form domain of $A$).

Using the scaling $\mathbf{x}\to Z\mathbf{x}$ one can readily see
that $H_{Z}^{a}$ (the Hamiltonian $H^{a}$ for a given constant $Z$)
is unitarily equivalent to $Z^{2}H_{Z=1}^{a}$. This is why we can
set, without loss of generality, $Z=1$, and this is what we do in
the remainder of the paper.

\subsection{The effective Hamiltonian}

The operator $-\Delta_{D}$ can be decomposed with respect to the
basis in $L^{2}((-a/2,a/2),\mbox{d}z)$ formed by the transversal
modes, \[
-\Delta_{D}=\bigoplus_{n=1}^{\infty}\left(-\Delta_{x,y}+E_{n}^{a}\right)\otimes\langle\chi_{n}^{a},\ \cdot\ \rangle\,\chi_{n}^{a},\]
with\[
E_{n}^{a}=\frac{n^{2}\pi^{2}}{a^{2}}\,,\ \,\chi_{n}^{a}(z)=\sqrt{\frac{2}{a}}\,\begin{cases}
\cos(n\pi z/a) & \text{ if n is odd}\\
\noalign{\smallskip}\sin(n\pi z/a) & \text{ if n is even}\end{cases}\,,\ n\in\mathbb{N.}\]
Here $-\Delta_{x,y}$ is the free Hamiltonian in $L^{2}(\mathbb{R}^{2},\mbox{d}x\mbox{d}y)$.
Put \[
P_{n}^{a}=1\otimes\langle\chi_{n}^{a},\ \cdot\ \rangle\chi_{n}^{a},\ \, n\in\mathbb{N}.\]
Using the projection on the lowest transversal mode we define the
effective Hamiltonian, \[
H_{\mathrm{eff}}^{a}=P_{1}^{a}H^{a}P_{1}^{a}.\]

This Hamiltonian may be regarded as an operator on $L^{2}(\mathbb{R}^{2})$,
\begin{equation}
H_{\mathrm{eff}}^{a}=-\Delta_{x,y}+E_{1}^{a}-V_{\mathrm{eff}}^{a}(\varrho)\label{eq:Heff_in_R2}\end{equation}
where the effective potential is defined by \begin{equation}
V_{\text{eff}}^{a}(\varrho)=\frac{2}{a}\int_{-a/2}^{a/2}\,\frac{\cos^{2}\left(\pi z/a\right)}{\sqrt{\varrho^{2}+z^{2}}}\,\mathrm{d}z\label{eq:Vaeff_def}\end{equation}
and $\varrho=\sqrt{x^{2}+y^{2}}$. Note that $0<V_{\text{eff}}^{a}(\varrho)<1/\varrho$
for all $\varrho>0$. Hence, if the Coulomb-like potential is $(-\Delta)$
form bounded with relative bound zero than the same is true for $V_{\text{eff}}^{a}$.
Thus the RHS in (\ref{eq:Heff_in_R2}) makes sense as a form sum and
$Q(H_{\text{eff}}^{a})=\mathcal{H}^{1}(\mathbb{R}^{2})$. Moreover,
\begin{equation}
-1+E_{1}^{a}\leq H_{C}+E_{1}^{a}\leq H_{\mathrm{eff}}^{a}.\label{eq:HC_leq_Haeff}\end{equation}
with $H_{C}$ being defined in (\ref{eq:H_C_def}) (also denoted by
$H(\infty)$ in the previous section). Let us also note that $\sigma_{\mathrm{ess}}(H_{\mathrm{eff}}^{a})=[\, E_{1}^{a},\infty)$.

It is even true that $V_{\mathrm{eff}}^{a}$ is $(-\Delta)$ bounded
with relative bound zero. In fact, recall that for any $\alpha>0$
there is $\beta$ such that\begin{equation}
\forall f\in\mathcal{H}^{2}(\mathbb{R}^{2}),\,\|f\|_{\infty}\leq\alpha\|\Delta f\|+\beta\|f\|,\label{eq:finf_leq_Hfreef}\end{equation}
with $\|\cdot\|$ being the $L^{2}$ norm, see \cite[Theorem~IX.28]{rs2}.
Moreover, one observes that\[
V_{\mathrm{eff}}^{a}(\varrho)=-\frac{4}{a}\ln(\varrho)+O(1)\ \,\mbox{as\ }\,\varrho\to0+\,,\]
and $V_{\mathrm{eff}}^{a}(\varrho)$ decays like $1/\varrho$ at infinity.
Hence $V_{\mathrm{eff}}^{a}$, regarded as a function on $\mathbb{R}^{2}$,
is square integrable at the origin and tends to zero at infinity.
It follows that for every $\epsilon>0$ there exists a decomposition
\begin{equation}
V_{\text{eff}}^{a}=V_{0}+V_{1},\ \mbox{with\ }\, V_{0}\in L^{\infty}(\mathbb{R}^{2}),\, V_{1}\in L^{2}(\mathbb{R}^{2}),\label{eq:Veff_decomp}\end{equation}
such that $\|V_{0}\|_{\infty}<\epsilon$. Combining (\ref{eq:finf_leq_Hfreef})
and (\ref{eq:Veff_decomp}) one finds that, for all $f\in\mathcal{H}^{2}(\mathbb{R}^{2})$,\[
\|V_{\text{eff}}^{a}f\|\leq\|V_{0}\|_{\infty}\|f\|+\|V_{1}\|\|f\|_{\infty}\leq\alpha\|V_{1}\|\|\Delta f\|+\left(\beta\,\|V_{1}\|+\|V_{0}\|_{\infty}\right)\|f\|.\]
This shows the relative boundedness and thus one can apply the Kato-Rellich
theorem. In particular, $\Dom H_{\mathrm{eff}}^{a}$ coincides with
$\Dom(-\Delta)=\mathcal{H}^{2}(\mathbb{R}^{2})$. Moreover, the existence
of decomposition (\ref{eq:Veff_decomp}) implies that $\sigma_{\mathrm{ess}}(H_{\mathrm{eff}}^{a})=[\, E_{1}^{a},\infty)$,
see \cite[Theorem~XIII.15]{rs4}.

\subsection{The limit of the effective Hamiltonian for small $a$ \label{sec:coul_eff}}

Here we show that the Hamiltonian $H_{\mathrm{eff}}^{a}-E_{1}^{a}$
converges to the two-dimensional hydrogenic Hamiltonian $H_{C}$ in
the norm resolvent sense as $a\to0+$.

\begin{lemma}\label{lem:free_coul_res} One has $\|(-\Delta+2)^{1/2}(H_{C}+2)^{-1/2}\|\leq C_{\mathrm{I}}$
where

\begin{equation}
C_{\mathrm{I}}=\frac{1}{8\pi^{2}}\!\left(\Gamma\!\left(\frac{1}{4}\right)^{\!4}+\sqrt{\Gamma\!\left(\frac{1}{4}\right)^{\!8}+64\pi^{4}}\,\right)\!.\label{eq:f_c_const}\end{equation}
\end{lemma}

\begin{proof} Put $L=(-\Delta+2)^{1/2}(H_{C}+2)^{-1/2}$. Then $L$
is bounded by the closed graph theorem but one can derive an upper
bound explicitly with the aid of the Kato inequality (\ref{eq:Kato_ineq}).
Since\[
\langle\psi,(-\Delta+2)^{-1/4}\varrho^{-1}(-\Delta+2)^{-1/4}\psi\rangle\leq\frac{\Gamma\!\left(\frac{1}{4}\right)^{4}}{4\pi^{2}}\,\|(-\Delta)^{1/4}(-\Delta+2)^{-1/4}\psi\|^{2}\leq\frac{\Gamma\!\left(\frac{1}{4}\right)^{4}}{4\pi^{2}}\,\|\psi\|^{2}\]
one has\[
L^{\dagger}L=1+(H_{C}+2)^{-1/2}\,\frac{1}{\varrho}\,(H_{C}+2)^{-1/2}\leq1+\frac{\Gamma\!\left(\frac{1}{4}\right)^{4}}{4\pi^{2}}\,(H_{C}+2)^{-1/2}(-\Delta+2)^{1/2}(H_{C}+2)^{-1/2}.\]
It follows that \[
\|L\psi\|^{2}\leq\|\psi\|^{2}+\frac{\Gamma\!\left(\frac{1}{4}\right)^{4}}{4\pi^{2}}\,\|(H_{C}+2)^{-1/2}\|\,\|\psi\|\,\|L\psi\|.\]
For $\|(H_{C}+2)^{-1/2}\|\leq1$ we get \[
\|L\|^{2}\leq1+\frac{\Gamma\!\left(\frac{1}{4}\right)^{4}}{4\pi^{2}}\,\|L\|.\]
Consequently, $\|L\|\leq C_{\mathrm{I}}$. \end{proof}

\begin{lemma}\label{lem:gen_dif_est} Suppose $W\in L^{1}(\mathbb{R}_{+},\mathrm{d}\varrho)$
and $0\leq W\leq1$. Put \begin{equation}
V^{a}(\varrho)=\frac{1}{\varrho}\left(1-W\!\left(\frac{\varrho}{a}\right)\right)\!,\ a>0.\label{eq:Va_def}\end{equation}
Then for any $a$, $0<a<1/2$, one has \begin{equation}
\begin{split} & \big\|(-\Delta+2)^{-1/2}\left(\varrho^{-1}-V^{a}\right)(-\Delta+2)^{-1/2}\big\|^{2}\\
\noalign{\smallskip} & \leq\,12a^{2}\ln^{2}(a)\left(\int_{\mathbb{R}_{+}}W(\varrho)\,\mathrm{d}\varrho\right)^{\!2}+32a^{2}\int_{\mathbb{R}_{+}}W(\varrho)\,\mathrm{d}\varrho.\end{split}
\label{eq:dif_est}\end{equation}
\end{lemma}

\begin{proof} Put \begin{equation}
T_{a}=(\varrho^{-1}-V^{a})^{1/2}\,(-\Delta+1)^{-1/2}.\label{eq:Ta_def}\end{equation}
Then \[
T_{a}^{\dagger}T_{a}={\textstyle (-\Delta+1)^{-1/2}\left(\varrho^{-1}-V^{a}\right)(-\Delta+1)^{-1/2}}\]
and $\|T_{a}^{\dagger}T_{a}\|=\|T_{a}T_{a}^{\dagger}\|$. Let us estimate
the Hilbert-Schmidt norm of $T_{a}T_{a}^{\dagger}$. The integral
kernel of $T_{a}T_{a}^{\dagger}$ is \[
\mathcal{K}(\mathbf{x}_{1},\mathbf{x}_{2})=\frac{1}{2\pi}\,\sqrt{\frac{1}{\varrho_{1}}\, W\!\left(\frac{\varrho_{1}}{a}\right)}\, K_{0}(|\mathbf{x}_{1}-\mathbf{x}_{2}|)\sqrt{\frac{1}{\varrho_{2}}\, W\!\left(\frac{\varrho_{2}}{a}\right)}\]
where $\mathbf{x}_{i}=\varrho_{i}(\cos\varphi_{i},\,\sin\varphi_{i})$.
Since the modified Bessel function $K_{0}$ is positive and strictly
decreasing on $\mathbb{R}_{+}$ we get \[
\|T_{a}T_{a}^{\dagger}\|_{\mathrm{HS}}^{\,2}\leq I(\mathbb{R}_{+}\times\mathbb{R}_{+})\]
where the symbol $I(M)$, $M\subset\mathbb{R}_{+}\times\mathbb{R}_{+}$
measurable, is defined by \[
I(M)=\int_{M}W\!\left(\frac{\varrho_{1}}{a}\right)K_{0}(|\varrho_{1}-\varrho_{2}|)^{2}\, W\!\left(\frac{\varrho_{2}}{a}\right)\,\mathrm{d}\varrho_{1}\mathrm{d}\varrho_{2}.\]

For $0<a<1/2$ and $|\varrho_{1}-\varrho_{2}|>a$ one has \cite{as}\begin{align*}
K_{0}(|\varrho_{1}-\varrho_{2}|) & <K_{0}(a)<\!\left(\!\ln\!\left(\frac{2}{a}\right)-\gamma\right)\! I_{0}(a)+\frac{1}{2}\, I_{0}(\sqrt{2}\, a)-\frac{1}{2}\\
 & <\left(\!\ln\!\left(\frac{2}{a}\right)-\gamma\right)\! I_{0}\!\left(\frac{1}{2}\right)\!+\frac{1}{2}\, I_{0}\!\left(\frac{1}{\sqrt{2}}\right)\!-\frac{1}{2}\,<-2\ln(a).\end{align*}
Consequently, \[
I(\left\lbrace |\varrho_{1}-\varrho_{2}|>a\right\rbrace )\leq4a^{2}\ln^{2}(a)\left(\int_{\mathbb{R}_{+}}W(\varrho)\,\mathrm{d}\varrho\right)^{\!2}.\]
 If $|\varrho_{1}-\varrho_{2}|<a<1/2$ then $K_{0}(a|\varrho_{1}-\varrho_{2}|)<-2\ln|\varrho_{1}-\varrho_{2}|$.
Moreover, for $0\leq W\leq1$,\[
\int_{|\varrho_{1}-\varrho_{2}|<1}W(\varrho_{1})\ln^{2}(|\varrho_{1}-\varrho_{2}|)\,\mathrm{d}\varrho_{1}\leq\int_{|v|<1}\ln^{2}|v|\,\mathrm{d}v=4.\]
It follows that \[
\begin{split}I(\left\lbrace |\varrho_{1}-\varrho_{2}|<a\right\rbrace ) & \leq4\,\int_{|\varrho_{1}-\varrho_{2}|<a}W\!\left(\frac{\varrho_{1}}{a}\right)\ln^{2}(|\varrho_{1}-\varrho_{2}|)\, W\!\left(\frac{\varrho_{2}}{a}\right)\,\mathrm{d}\varrho_{1}\mathrm{d}\varrho_{2}\\
 & \leq\,8a^{2}\int_{|\varrho_{1}-\varrho_{2}|<1}W(\varrho_{1})\left(\ln^{2}(a)+\ln^{2}|\varrho_{1}-\varrho_{2}|\right)W(\varrho_{2})\,\mathrm{d}\varrho_{1}\mathrm{d}\varrho_{2}\\
 & \leq\,8a^{2}\ln^{2}(a)\left(\int_{\mathbb{R}_{+}}W(\varrho)\,\mathrm{d}\varrho\right)^{2}+32a^{2}\int_{\mathbb{R}_{+}}W(\varrho)\,\mathrm{d}\varrho.\end{split}
\]

We conclude that \[
\|T_{a}T_{a}^{\dagger}\|_{\mathrm{HS}}^{\,2}\,\leq12a^{2}\ln^{2}(a)\left(\int_{\mathbb{R}_{+}}W(\varrho)\,\mathrm{d}\varrho\right)^{2}+32a^{2}\int_{\mathbb{R}_{+}}W(\varrho)\,\mathrm{d}\varrho.\]
By the functional calculus, $\|(-\Delta+2)^{-1/2}(-\Delta+{\textstyle 1})^{1/2}\|=1$,
and this completes the proof. \end{proof}

\begin{lemma}\label{lem:lower_bound} Suppose $W\in L^{1}(\mathbb{R}_{+},\mathrm{d}\varrho)$
and $W(\varrho)\geq0$. Let $V^{a}(\varrho)$ be defined as in (\ref{eq:Va_def}).
Then \[
\big\|(-\Delta+2)^{-1/2}\left(\varrho^{-1}-V^{a}\right)(-\Delta+2)^{-1/2}\big\|\geq\frac{1}{2}\!\left(\int_{0}^{R}W(\varrho)\mathrm{d}\varrho\right)a\ln\!\left(\frac{1}{aR}\right)\]
whenever $R>1$ and $a>0$. \end{lemma}

\begin{proof} We again use definition (\ref{eq:Ta_def}). Chose $R>1$
and truncate $\tilde{W}(\varrho)=W(\varrho)\vartheta(R-\varrho)$
where $\vartheta(x)$ is the Heaviside step function (the characteristic
function of the positive halfline). If $f\in L^{2}(\mathbb{R}^{2},\mathrm{d}\mathbf{x})$,
$f\neq0$, then $\|T_{a}T_{a}^{\dagger}\|\geq\langle f,T_{a}T_{a}^{\dagger}f\rangle/\|f\|^{2}$.
We choose \[
f(\mathbf{x})=\left[\frac{1}{|\mathbf{x}|}\,\tilde{W}\!\left(\frac{|\mathbf{x}|}{a}\right)\right]^{\!1/2}.\]
Then \[
\|f\|^{2}=2\pi a\int_{0}^{R}W(\varrho)\mathrm{d}\varrho\]
and \[
\begin{split}\langle f,T_{a}T_{a}^{\dagger}f\rangle= & \frac{1}{2\pi}\int_{\mathbb{R}^{2}}\int_{\mathbb{R}^{2}}K_{0}(|\mathbf{x}_{1}-\mathbf{x}_{2}|)\frac{1}{|\mathbf{x}_{1}|}\tilde{W}\!\left(\frac{|\mathbf{x}_{1}|}{a}\right)\frac{1}{|\mathbf{x}_{2}|}\tilde{W}\!\left(\frac{|\mathbf{x}_{2}|}{a}\right)\mathrm{d}\mathbf{x}_{1}\mathrm{d}\mathbf{x}_{2}\\
= & \frac{1}{2\pi}\int_{\mathbb{R}_{+}\times S^{1}}\int_{\mathbb{R}_{+}\times S^{1}}K_{0}\!\left(\big(\varrho_{1}^{2}+\varrho_{2}^{2}-2\varrho_{1}\varrho_{2}\cos(\varphi_{1}-\varphi_{2})\big)^{1/2}\right)\\
 & \times\,\tilde{W}\!\left(\frac{\varrho_{1}}{a}\right)\tilde{W}\!\left(\frac{\varrho_{2}}{a}\right)\mathrm{d}\varrho_{1}\mathrm{d}\varphi_{1}\mathrm{d}\varrho_{2}\mathrm{d}\varphi_{2}.\end{split}
\]

Recall that, by formula 11.4.44 in \cite{as}, \[
K_{0}\!\left(\big(\varrho_{1}^{2}+\varrho_{2}^{2}-2\varrho_{1}\varrho_{2}\cos\varphi\big)^{1/2}\right)=\int_{0}^{\infty}J_{0}\!\left(\big(\varrho_{1}^{2}+\varrho_{2}^{2}-2\varrho_{1}\varrho_{2}\cos\varphi\big)^{1/2}t\right)\frac{t}{t^{2}+1}\,\mathrm{d}t.\]
Integrating Graf's addition formula for Bessel functions we obtain
\[
\frac{1}{2\pi}\int_{0}^{2\pi}J_{0}\!\left(\big(\varrho_{1}^{2}+\varrho_{2}^{2}-2\varrho_{1}\varrho_{2}\cos\varphi\big)^{1/2}t\right)\mathrm{d}\varphi=J_{0}(\varrho_{1}t)J_{0}(\varrho_{2}t).\]
For \[
\int_{0}^{\infty}J_{0}(\varrho_{1}t)J_{0}(\varrho_{2}t)\,\frac{t}{t^{2}+1}\mathrm{\, d}t=I_{0}(\varrho_{<})K_{0}(\varrho_{>})\]
we conclude that \[
\frac{1}{2\pi}\int_{0}^{2\pi}K_{0}\!\left(\big(\varrho_{1}^{2}+\varrho_{2}^{2}-2\varrho_{1}\varrho_{2}\cos\varphi\big)^{1/2}\right)\mathrm{d}\varphi=I_{0}(\varrho_{<})K_{0}(\varrho_{>}).\]
Also recall that $I_{0}(\varrho)\geq1$ and $K_{0}(\varrho)\geq\ln(2/\varrho)-\gamma\geq\ln(1/\varrho)$.

For any $a>0$ we get \[
\begin{split}\langle f,T_{a}T_{a}^{\dagger}f\rangle & =2\pi\int_{\mathbb{R}_{+}}\int_{\mathbb{R}_{+}}I_{0}(\varrho_{<})K_{0}(\varrho_{>})\tilde{W}\!\left(\frac{\varrho_{1}}{a}\right)\tilde{W}\!\left(\frac{\varrho_{2}}{a}\right)\mathrm{d}\varrho_{1}\mathrm{d}\varrho_{2}\\
 & \geq4\pi a^{2}\int_{0}^{R}\left(\int_{\varrho_{2}}^{R}\ln\!\left(\frac{1}{a\varrho_{1}}\right)W(\varrho_{1})\mathrm{d}\varrho_{1}\right)W(\varrho_{2})\mathrm{d}\varrho_{2}\\
 & \geq2\pi a^{2}\,\ln\!\left(\frac{1}{aR}\right)\left(\int_{0}^{R}W(\varrho)\mathrm{d}\varrho\right)^{\!2}.\end{split}
\]
Finally note that $\|(-\Delta+1)^{-1/2}(-\Delta+{\textstyle 2})^{1/2}\|=\sqrt{2}$.
The lemma follows. \end{proof}

\begin{remark} \label{rem:Vaeff_W} Note that the effective potential
has the scaling property \[
V_{\mathrm{eff}}^{a}(\varrho)=\frac{1}{a}\, V_{\mathrm{eff}}^{1}\!\left(\frac{\varrho}{a}\right).\]
Recall also that $0<V_{\mathrm{eff}}^{a}(\varrho)<1/\varrho$. If
we put\begin{equation}
W(\varrho)=1-\varrho V_{\mathrm{eff}}^{1}(\varrho)\label{eq:W_Vaeff}\end{equation}
then $0<W(\varrho)<1$ and\[
\frac{1}{\varrho}\left(1-W\!\left(\frac{\varrho}{a}\right)\right)=V_{\mathrm{eff}}^{a}(\varrho).\]
Moreover, from (\ref{eq:Vaeff_def}) one gets\[
W(\varrho)=2\int_{-1/2}^{1/2}\left(1-\frac{\varrho}{\sqrt{\varrho^{2}+z^{2}}}\right)\!\cos^{2}(\pi z)\,\mbox{d}z\leq\frac{1}{\varrho^{2}}\int_{-1/2}^{1/2}z^{2}\cos^{2}(\pi z)\,\mbox{d}z.\]
Hence $W\in L^{2}(\mathbb{R}_{+},\mbox{d}\varrho)$. Thus all assumptions
of Lemmas~\ref{lem:gen_dif_est} and \ref{lem:lower_bound} are fulfilled.
In the course of the proofs of these lemmas we have shown that there
exist constants $0<C_{1}<C_{2}$ such that for all sufficiently small
$a>0$, \[
C_{1}a\,|\!\ln a|<\big\|(-\Delta+1)^{-1/2}(\varrho^{-1}-V_{\mathrm{eff}}^{a})(-\Delta+1)^{-1/2}\big\|<C_{2}a\,|\!\ln a|.\]
\end{remark}

Further we need an estimate formulated in the following lemma which
is easy to see and is in fact a standard result (see for instance,
\cite[Chp.~XI]{rs3}).

\begin{lemma} \label{thm:symm_resolv_eq} Assume that $A$ is semibounded,
$A^{-1}$ exists and is bounded, $C$ is self-adjoint and $A$ form
bounded. If\[
\alpha=\||C|^{1/2}|A|^{-1/2}\|<1\]
then $(A+C)^{-1}$ exists, is bounded and\[
\|(A+C)^{-1}-A^{-1}\|\leq\frac{\alpha^{2}\|A^{-1}\|}{1-\alpha^{2}}\,.\]
 \end{lemma}

\begin{proposition}\label{theo:coul_conv} For every $\xi\in\Res(H_{C})\cap\mathbb{R}$
there exists $a_{0}(\xi)>0$ such that for all $a$, $0<a<a_{0}(\xi)$,
one has $\xi\in\Res(H_{\mathrm{eff}}^{a}-E_{1}^{a})$ and \[
\big\|(H_{\mathrm{eff}}^{a}-E_{1}^{a}-\xi)^{-1}-(H_{C}-\xi)^{-1}\big\|\leq\frac{2C_{\mathrm{I}}^{\,2}C_{\mathrm{II}}}{d_{C}(\xi)}\,\max\!\left\lbrace 1\,,\,\frac{2}{d_{C}(\xi)}\right\rbrace \, a\,|\!\ln a|\]
where $d_{C}(\xi)=\mathrm{dist}(\xi,\sigma(H_{C}))$, $C_{\mathrm{I}}$
is given in (\ref{eq:f_c_const}) and \begin{equation}
C_{\mathrm{II}}=\frac{\sqrt{3}}{2}\left(1-\frac{4}{\pi^{2}}\right)\sqrt{1+\frac{32\pi^{2}}{3(\pi^{2}-4)\ln^{2}(2)}}\,.\label{eq:CII}\end{equation}
 \end{proposition}

\begin{proof} Note that $\xi<0$. One can apply Lemma~\ref{thm:symm_resolv_eq}
with $A=H_{C}-\xi$, $C=\varrho^{-1}-V_{\text{eff}}^{a}\,$. Then\[
A+C=H_{\mathrm{eff}}^{a}-E_{1}^{a}-\xi,\]
$\|A^{-1}\|=1/d_{C}(\xi)$, and one has\begin{align*}
 & \alpha^{2}=\||C|^{1/2}|A|^{-1/2}\|^{2}=\big\||H_{C}-\xi|^{-1/2}\left(\varrho^{-1}-V_{\text{eff}}^{a}\right)|H_{C}-\xi|^{-1/2}\big\|\\
\noalign{\smallskip} & \leq\,\big\|(H_{C}+2)^{1/2}|H_{C}-\xi|^{-1/2}\big\|^{2}\,\big\|(-\Delta+2)^{1/2}(H_{C}+2)^{-1/2}\big\|^{2}\\
\noalign{\smallskip} & \hspace{1em}\times\,\big\|(-\Delta+2)^{-1/2}\left(\varrho^{-1}-V_{\text{eff}}^{a}\right)(-\Delta+2)^{-1/2}\big\|\end{align*}
By Lemma~\ref{lem:free_coul_res}, $\|(-\Delta+2)^{1/2}(H_{C}+2)^{-1/2}\|\leq C_{\mathrm{I}}$.
Furthermore,\[
\big\|(H_{C}+2)^{1/2}|H_{C}-\xi|^{-1/2}\big\|^{2}=\sup_{x\in\sigma(H_{C})}\frac{x+2}{|x-\xi|}\leq\max\left\lbrace 1\,,\,\frac{2}{d_{C}(\xi)}\right\rbrace .\]
Finally, according to Remark~\ref{rem:Vaeff_W}, for the same $W$
as given in (\ref{eq:W_Vaeff}) one has $V^{a}=V_{\mathrm{eff}}^{a}$.
In that case,\[
\int_{0}^{\infty}W(\varrho)\mbox{d}\varrho=2\int_{-1/2}^{1/2}\left[\int_{0}^{\infty}\left(1-\frac{\varrho}{\sqrt{\varrho^{2}+z^{2}}}\right)\!\mbox{d}\varrho\right]\!\cos^{2}(\pi z)\,\mbox{d}z=\frac{1}{4}-\frac{1}{\pi^{2}}\,.\]
Suppose $0<a<1/2$. Recalling Lemma~\ref{lem:gen_dif_est} one derives
the estimate\[
\big\|(-\Delta+2)^{-1/2}\left(\varrho^{-1}-V^{a}\right)(-\Delta+2)^{-1/2}\big\|\leq C_{\mathrm{II}}\, a\,|\!\ln a|,\]
with $C_{\mathrm{II}}$ given in (\ref{eq:CII}). Hence \[
\alpha^{2}\leq C_{\mathrm{I}}^{\,2}C_{\mathrm{II}}\,\max\!\left\lbrace 1\,,\,\frac{2}{d_{C}(\xi)}\right\rbrace \, a\,|\!\ln a|.\]

Now it is clear that for any $\xi\in\Res(H_{C})\cap\mathbb{R}$ one
can find $a_{0}(\xi)>0$ such that for all $a$, $0<a<a_{0}(\xi)$,
one has $\alpha^{2}\leq1/2$. Then $(H_{\mathrm{eff}}^{a}-E_{1}^{a}-\xi)^{-1}$
exists, is bounded and \[
\big\|(H_{\mathrm{eff}}^{a}-E_{1}^{a}-\xi)^{-1}-(H_{C}-\xi)^{-1}\big\|\leq\frac{\alpha^{2}}{d_{C}(\xi)(1-\alpha^{2})}\leq\frac{2\alpha^{2}}{d_{C}(\xi)}\,.\]
This shows the theorem. \end{proof}

\subsection{Approximation of the full Hamiltonian by the effective Hamiltonian
\label{sec:eff_aprox}}

In this section we show that the effective Hamiltonian $H_{\text{eff}}^{a}$
tends to the full Hamiltonian $H^{a}$ in the norm resolvent sense
as $a\to0+$. To this end, we decompose the Hilbert space $L^{2}(\Omega_{a})$
into an orthogonal sum determined by the projection onto the first
transversal mode and the projection on all remaining higher-order
modes.

To simplify notation we write $P^{a}$ instead of $P_{1}^{a}$. In
this subsection we denote the Coulomb potential $-1/r$ in the slab
as $V$. Put\[
Q^{a}=1-P^{a},\ H_{\bot}^{a}=Q^{a}H^{a}Q^{a},\ R_{\bot}^{a}(\xi)=(H_{\bot}^{a}-\xi)^{-1}.\]
Furthermore, we denote\[
\mathscr W^{a}(\xi)=P^{a}VQ^{a}R_{\bot}^{a}(\xi)Q^{a}VP^{a},\ R_{\text{eff}}^{\mathscr{W}}(\xi)=\left(H_{\text{eff}}^{a}-\mathscr W^{a}(\xi)-\xi\right)^{-1}.\]
If convenient, $\mathscr W^{a}$ may be regarded as an operator in
$L^{2}(\mathbb{R}^{2})$. During some manipulations the dependence
of operators on the spectral parameter $\xi$ will not be indicated
explicitly.

With respect to the decomposition $L^{2}(\Omega_{a})=\Ran P^{a}\oplus\Ran Q^{a}$
one can write \[
H^{a}=\begin{pmatrix}H_{\text{eff}}^{a} & P^{a}H^{a}Q^{a}\\
Q^{a}H^{a}P^{a} & H_{\bot}^{a}\end{pmatrix}.\]
As is well known and easy to verify, if $A$, $B$, $C$, $D$ are
bounded operators between appropriate Hilbert spaces then the following
formula for inversion of the operator matrix,\[
\begin{pmatrix}A & B\\
C & D\end{pmatrix}^{\!-1}=\begin{pmatrix}W^{-1} & -W^{-1}BD^{-1}\\
-D^{-1}CW^{-1} & D^{-1}+D^{-1}CW^{-1}BD^{-1}\end{pmatrix},\ W=A-BD^{-1}C,\]
holds true provided $D^{-1}$ and $W^{-1}$ exist and are bounded.
This way one obtains a formula for the resolvent of $H^{a},$ provided
$R_{\text{eff}}^{\mathscr{W}}$ and $R_{\bot}^{a}$ do exist, namely
\begin{equation}
(H^{a}-\xi)^{-1}=\begin{pmatrix}R_{\text{eff}}^{\mathscr{W}} & -R_{\text{eff}}^{\mathscr{W}}P^{a}VQ^{a}R_{\bot}^{a}\\
-R_{\bot}^{a}Q^{a}VP^{a}R_{\text{eff}}^{\mathscr{W}} & R_{\bot}^{a}+R_{\bot}^{a}Q^{a}VP^{a}R_{\text{eff}}^{\mathscr{W}}P^{a}VQ^{a}R_{\bot}^{a}\end{pmatrix}.\label{eq:Feshbach}\end{equation}
Notice that $P^{a}H^{a}Q^{a}=P^{a}VQ^{a},\ Q^{a}H^{a}P^{a}=Q^{a}VP^{a}$.

\begin{proposition} \label{theo:fesh1} If $0<a<3\pi/8$, $\xi<E_{1}^{a}$
and $\xi\notin\sigma(H_{\mathrm{eff}}^{a}-\mathscr{W}^{a}(\xi))$
then $\xi\in\Res(H^{a})$ and \begin{equation}
\big\|(H^{a}-\xi)^{-1}-R_{\mathrm{eff}}^{\mathscr{W}}(\xi)\oplus0\big\|\leq\,\frac{8a}{3\pi d_{\mathrm{eff}}^{\mathscr{W}}(\xi)}\!\left(1+\frac{8a}{3\pi}\right)+\frac{2a^{2}}{3\pi^{2}}\label{eq:fesh1_res_estim}\end{equation}
where \[
d_{\mathrm{eff}}^{\mathscr{W}}(\xi)=\mathrm{dist}\big(\xi,\sigma(H_{\mathrm{eff}}^{a}-\mathscr{W}^{a}(\xi))\big).\]
\end{proposition}

\begin{proof} This proof is inspired by the proof of Theorem~3.1
in \cite{bd}. Note that\[
\Big\|\begin{pmatrix}0 & A\\
A^{\dagger} & 0\end{pmatrix}\Big\|^{2}=\Big\|\begin{pmatrix}AA^{\dagger} & 0\\
0 & A^{\dagger}A\end{pmatrix}\Big\|=\|AA^{\dagger}\|=\|A\|^{2}.\]
Thus from formula (\ref{eq:Feshbach}) one derives the estimate\begin{align}
\|(H^{a}-\xi)^{-1}-R_{\text{eff}}^{\mathscr{W}}(\xi)\oplus0\| & \leq\|R_{\text{eff}}^{\mathscr{W}}P^{a}VQ^{a}R_{\bot}^{a}\|+\|R_{\bot}^{a}Q^{a}VP^{a}R_{\text{eff}}^{\mathscr{W}}P^{a}VQ^{a}R_{\bot}^{a}\|+\|R_{\bot}^{a}\|\nonumber \\
 & \leq\frac{1}{d_{\mathrm{eff}}^{\mathscr{W}}}\,\|VQ^{a}R_{\bot}^{a}\|\left(1+\|VQ^{a}R_{\bot}^{a}\|\right)+\|R_{\bot}^{a}\|.\label{eq:fesh1_intermed}\end{align}
To complete the proof one has to show that $\xi\in\Res(H_{\bot}^{a})$
and estimate $\|R_{\bot}^{a}\|$ and $\|VQ^{a}R_{\bot}^{a}\|$. 

Let us denote (in this proof)\[
T_{\bot}=Q^{a}(-\Delta_{D})Q^{a},\ R_{0}=(T_{\bot}-\xi)^{-1}.\]
Since $\xi<E_{1}^{a}=\pi^{2}/a^{2}$ and \[
T_{\bot}=Q^{a}\!\left(-\Delta_{x,y}\otimes1\right)Q^{a}+Q^{a}\!\left(-1\otimes\partial_{z}^{2}\right)\! Q^{a}\geq E_{2}^{a}=4\pi^{2}/a^{2}\]
one has \[
0\leq R_{0}\leq(E_{2}^{a}-E_{1}^{a})^{-1}=\frac{a^{2}}{3\pi^{2}}\,,\ \,\xi R_{0}\leq\frac{1}{3}\,.\]
Further let us estimate $\big\|VQ^{a}R_{0}^{\,1/2}\big\|=\big\|R_{0}^{\,1/2}Q^{a}V^{2}Q^{a}R_{0}^{\,1/2}\big\|^{1/2}$.
By the Hardy inequality (\ref{eq:Hardy}),\[
R_{0}^{\,1/2}Q^{a}V^{2}Q^{a}R_{0}^{\,1/2}\leq4R_{0}^{\,1/2}T_{\perp}R_{0}^{\,1/2}=4(Q^{a}+\xi R_{0})\leq\frac{16}{3}\,.\]
Hence \[
\big\|VQ^{a}R_{0}^{\,1/2}\big\|\leq\frac{4}{\sqrt{3}}\,.\]
Moreover,\[
\left(R_{0}^{\,1/2}Q^{a}VQ^{a}R_{0}^{\,1/2}\right)^{2}=R_{0}^{\,1/2}Q^{a}VQ^{a}R_{0}Q^{a}VQ^{a}R_{0}^{\,1/2}\leq\frac{a^{2}}{3\pi^{2}}\, R_{0}^{\,1/2}Q^{a}V^{2}Q^{a}R_{0}^{\,1/2}\]
and so \begin{equation}
\big\|R_{0}^{\,1/2}Q^{a}VQ^{a}R_{0}^{\,1/2}\big\|\leq\frac{4a}{3\pi}\,.\label{eq:fesh1_aux1}\end{equation}

Put $a_{H}=3\pi/4$. If $a<a_{H}$ then, by (\ref{eq:fesh1_aux1})
and the resolvent formula\begin{equation}
R_{\bot}^{a}(\xi)=(T_{\bot}+Q^{a}VQ^{a}-\xi)^{-1}=R_{0}^{\,1/2}\left(1+R_{0}^{\,1/2}Q^{a}VQ^{a}R_{0}^{\,1/2}\right)^{-1}R_{0}^{\,1/2},\label{eq:Rperp_R0}\end{equation}
one has $\xi\in\Res(H_{\perp}^{a})$ and the resolvent $R_{\bot}^{a}(\xi)$
is positive. Moreover, \[
\|R_{\bot}^{a}(\xi)\|\leq\frac{\|R_{0}\|}{1-\big\|R_{0}^{\,1/2}Q^{a}VQ^{a}R_{0}^{\,1/2}\big\|}\,.\]
For $a<a_{H}/2$, \begin{equation}
\|R_{\bot}^{a}\|\leq2\|R_{0}\|\leq\frac{2a^{2}}{3\pi^{2}}\,.\label{eq:res_bot_bound}\end{equation}
From (\ref{eq:Rperp_R0}) it follows that \[
\|VQ^{a}R_{\bot}^{a}Q^{a}V\|\leq\frac{\|VQ^{a}R_{0}^{\,1/2}\|^{2}}{1-\big\|R_{0}^{\,1/2}Q^{a}VQ^{a}R_{0}^{\,1/2}\big\|}\]
and this implies, again for $a<a_{H}/2$, \begin{equation}
\|VQ^{a}R_{\bot}^{a}\|\leq\|VQ^{a}(R_{\bot}^{a})^{1/2}\|\,\|(R_{\bot}^{a})^{1/2}\|\leq\frac{\|VQ^{a}R_{0}^{\,1/2}\|\,\|R_{\bot}^{a}\|^{1/2}}{\left(1-\big\|R_{0}^{\,1/2}Q^{a}VQ^{a}R_{0}^{\,1/2}\big\|\right)^{\!1/2}}\leq\frac{8a}{3\pi}\,.\label{eq:VQRperp_estim}\end{equation}

Finally we conclude that (\ref{eq:fesh1_intermed}), (\ref{eq:res_bot_bound})
and (\ref{eq:VQRperp_estim}) imply (\ref{eq:fesh1_res_estim}).

\end{proof}

\begin{lemma} \label{lem:1_fesh_2} If $0<a<3\pi/8$ and $\xi<E_{1}^{a}$
then $\mathscr{W}^{a}(\xi)$ is positive and \begin{equation}
\big\|(-\Delta+2)^{-1/2}\mathscr{W}^{a}(\xi)(-\Delta+2)^{-1/2}\big\|\leq\frac{\Gamma(1/4)^{4}}{6\sqrt{2}\,\pi^{3}}\, a\label{eq:free_Wa_free_bound}\end{equation}
where $-\Delta$ is the free Hamiltonian in $L^{2}(\mathbb{R}^{2})$.
\end{lemma}

\begin{proof} In the course of proof of Proposition~\ref{theo:fesh1},
when discussing formula (\ref{eq:Rperp_R0}), it is shown that if
$0<a<3\pi/8$ and $\xi<E_{1}^{a}$ then $R_{\perp}^{a}(\xi)$ is positive
and so is $\mathscr{W}^{a}(\xi)$. Using (\ref{eq:res_bot_bound})
we get \[
\begin{split}0\leq\mathscr{W}^{a} & =P^{a}VQ^{a}R_{\bot}^{a}Q^{a}VP^{a}\leq\frac{2a^{2}}{3\pi^{2}}\, P^{a}V^{2}P^{a}\\
 & =\frac{8a}{3\pi^{2}}\int_{0}^{a/2}\frac{\cos^{2}(\pi z/a)}{\varrho^{2}+z^{2}}\,\mathrm{d}z\leq\frac{8a}{3\pi^{2}}\int_{0}^{\infty}\frac{1}{\varrho^{2}+z^{2}}\,\mathrm{d}z=\,\frac{4a}{3\pi\varrho}\,.\end{split}
\]
Recalling the Kato inequality (\ref{eq:Kato_ineq}) one finds that
\begin{align*}
(-\Delta+2)^{-1/2}\mathscr{W}^{a}(-\Delta+2)^{-1/2} & \leq\frac{4a}{3\pi}\,(-\Delta+2)^{-1/2}\,\frac{1}{\varrho}\,(-\Delta+2)^{-1/2}\\
 & \leq\frac{\Gamma(1/4)^{4}a}{3\pi^{3}}\,\sqrt{-\Delta}\,(-\Delta+2)^{-1}.\end{align*}
The lemma readily follows. \end{proof}

\begin{lemma}\label{lem:2_fesh_2} If $\mu\leq E_{1}^{a}-2$ then
\begin{equation}
\|(-\Delta+2)^{1/2}(H_{\mathrm{eff}}^{a}-\mu)^{-1/2}\|\leq C_{\mathrm{I}},\label{eq:free_Haeff_CI}\end{equation}
with $C_{\mathrm{I}}$ being given in (\ref{eq:f_c_const}). \end{lemma}

\begin{proof} Since $0\leq V_{\mathrm{eff}}^{a}(\varrho)\leq1/\varrho$
the Kato inequality (\ref{eq:Kato_ineq}) implies\[
V_{\mathrm{eff}}^{a}(\varrho)\leq\frac{\Gamma(1/4)^{4}}{4\pi^{2}}\,\sqrt{-\Delta}\,.\]
Note that, in virtue of (\ref{eq:HC_leq_Haeff}), $0<(H_{\mathrm{eff}}^{a}-\mu)^{-1}\leq1$
if $\mu\leq E_{1}^{a}-2$. Now, to show (\ref{eq:free_Haeff_CI}),
one can repeat word by word the proof of Lemma \ref{lem:free_coul_res}
while replacing $1/\varrho$ by $V_{\mathrm{eff}}^{a}(\varrho)$ and
$H_{C}+2$ by $H_{\mathrm{eff}}^{a}-\mu$. \end{proof}

\begin{proposition}\label{theo:fesh2} Suppose that $\xi\in\Res(H_{\mathrm{eff}}^{a})\cap\mathbb{R}$.
If \begin{equation}
a<\frac{1}{2C_{\mathrm{III}}}\,\min\left\lbrace 1,\ \frac{d_{\mathrm{eff}}(\xi)}{2}\right\rbrace \label{eq:fesh2_cond}\end{equation}
where $d_{\mathrm{eff}}(\xi)=\mathrm{dist}(\xi,\ \sigma(H_{\mathrm{eff}}^{a}))$,\begin{equation}
C_{\mathrm{III}}=C_{\mathrm{I}}^{\,2}\,\frac{\Gamma(1/4)^{4}}{6\sqrt{2}\,\pi^{3}}\,,\label{eq:c_fesh2}\end{equation}
and $C_{\mathrm{I}}$ is defined in (\ref{eq:f_c_const}) then $\xi\notin\sigma(H_{\mathrm{eff}}^{a}-\mathscr{W}^{a}(\xi))$
and \begin{equation}
\|R_{\mathrm{eff}}^{\mathscr{W}}(\xi)-(H_{\mathrm{eff}}^{a}-\xi)^{-1}\|\leq\frac{2C_{\mathrm{III}}}{d_{\mathrm{eff}}(\xi)}\,\max\left\lbrace 1\,,\,\frac{2}{d_{\mathrm{eff}}(\xi)}\right\rbrace a.\label{eq:RWeff_ResHaeff_estim}\end{equation}
\end{proposition}

\begin{proof} Apply Lemma~\ref{thm:symm_resolv_eq} with $A=H_{\mathrm{eff}}^{a}-\xi$,
$C=-\mathscr{W}^{a}(\xi)$. Since $\xi\in\Res(H_{\mathrm{eff}}^{a})\cap\mathbb{R}$
one has $\xi<E_{1}^{a}$. Furthermore, in view of Lemma~\ref{lem:1_fesh_2},
one observes that $\mathscr{W}^{a}(\xi)$ is positive provided $0<a<3\pi/8$.
By Lemma~\ref{thm:symm_resolv_eq}, if\[
\alpha=\big\|\mathscr{W}^{a}(\xi)^{1/2}|H_{\mathrm{eff}}^{a}-\xi|^{-1/2}\big\|<1\]
then $\xi\notin\sigma(H_{\mathrm{eff}}^{a}-\mathscr{W}^{a}(\xi))$
and \begin{equation}
\|R_{\mathrm{eff}}^{\mathscr{W}}(\xi)-(H_{\mathrm{eff}}^{a}-\xi)^{-1}\|\leq\frac{\alpha^{2}}{d_{\text{eff}}(\xi)(1-\alpha^{2})}\,.\label{eq:res_est_fesh2}\end{equation}

If $\mu<E_{1}^{a}-1$ then, according to (\ref{eq:HC_leq_Haeff}),
$H_{\mathrm{eff}}^{a}-\mu$ is positive. One has\begin{align*}
\alpha^{2} & =\big\||H_{\mathrm{eff}}^{a}-\xi|^{-1/2}\mathscr{W}^{a}|H_{\mathrm{eff}}^{a}-\xi|^{-1/2}\big\|\\
\noalign{\smallskip} & \leq\big\|(-\Delta+2)^{-1/2}\mathscr{W}^{a}(-\Delta+2)^{-1/2}\big\|\,\|(-\Delta+2)^{1/2}(H_{\mathrm{eff}}^{a}-\mu)^{-1/2}\|^{2}\\
\noalign{\smallskip} & \hspace{1em}\,\times\,\big\|(H_{\mathrm{eff}}^{a}-\mu)^{1/2}|H_{\mathrm{eff}}^{a}-\xi|^{-1/2}\big\|^{2}.\end{align*}
Observe that\begin{equation}
\big\|(H_{\mathrm{eff}}^{a}-\mu)^{1/2}|H_{\mathrm{eff}}^{a}-\xi|^{-1/2}\big\|^{2}=\sup_{x\in\sigma(H_{\mathrm{eff}}^{a})}\frac{x-\mu}{|x-\xi|}\leq\max\left\lbrace 1\,,\,\frac{E_{1}^{a}-\mu}{d_{\text{eff}}(\xi)}\right\rbrace .\label{eq:Haeff_mu_xi_bound}\end{equation}
Set $\mu=E_{1}^{a}-2$. Then (\ref{eq:Haeff_mu_xi_bound}) jointly
with (\ref{eq:free_Wa_free_bound}), (\ref{eq:free_Haeff_CI}) imply\[
\alpha^{2}\leq C_{\mathrm{III}}\,\max\left\lbrace 1\,,\,\frac{2}{d_{\text{eff}}(\xi)}\right\rbrace \, a.\]
If condition (\ref{eq:fesh2_cond}) is satisfied then $\alpha^{2}<1/2$
and (\ref{eq:RWeff_ResHaeff_estim}) follows from (\ref{eq:res_est_fesh2}).
\end{proof}

\begin{remark} Under the assumptions of Proposition~\ref{theo:fesh2},
$\alpha$ in (\ref{eq:res_est_fesh2}) fulfills $\alpha^{2}<1/2$
and so\[
\|R_{\mathrm{eff}}^{\mathscr{W}}(\xi)-(H_{\mathrm{eff}}^{a}-\xi)^{-1}\|\leq\|(H_{\mathrm{eff}}^{a}-\xi)^{-1}\|\]
whence \[
\|R_{\mathrm{eff}}^{\mathscr{W}}(\xi)\|\leq2\|(H_{\mathrm{eff}}^{a}-\xi)^{-1}\|.\]
This means that \begin{equation}
\frac{1}{d_{\mathrm{eff}}^{\mathscr{W}}(\xi)}\leq\frac{2}{d_{\mathrm{eff}}(\xi)}\,.\label{eq:dist_ineq}\end{equation}
Similarly, under the assumptions of Theorem~\ref{theo:coul_conv},
\begin{equation}
\frac{1}{d_{\mathrm{eff}}(\xi+E_{1}^{a})}\leq\frac{2}{d_{C}(\xi)}.\label{eq:dist_ineq_coul}\end{equation}
\end{remark}

\begin{proposition} \label{theo:fesh3} Assume that $\xi\in\Res(H_{\mathrm{eff}}^{a})\cap\mathbb{R}$
and $a>0$ fulfills (\ref{eq:fesh2_cond}). Then $\xi\in\Res(H^{a})$
and \[
\big\|(H^{a}-\xi)^{-1}-(H_{\mathrm{eff}}^{a}-\xi)^{-1}\oplus0\big\|\leq\left(\frac{8}{\pi}+\max\left\lbrace 1\,,\,\frac{2}{d_{\mathrm{eff}}(\xi)}\right\rbrace C_{\mathrm{III}}\right)\frac{2a}{d_{\mathrm{eff}}(\xi)}+\frac{2a^{2}}{3\pi^{2}}\]
where $C_{\mathrm{III}}$ is given in (\ref{eq:c_fesh2}). \end{proposition}

\begin{proof} If $\xi\in\Res(H_{\mathrm{eff}}^{a})\cap\mathbb{R}$
then $\xi<E_{1}^{a}$. Furthermore, from Proposition~\ref{theo:fesh2}
it follows that $\xi\notin\sigma(H_{\mathrm{eff}}^{a}-\mathscr{W}^{a}(\xi))$.
Observe also that, by the fact that $R_{\bot}^{a}(\xi)\geq0$ for
any $\xi<E_{1}^{a}$, one has $H_{\bot}^{a}>E_{1}^{a}$. Moreover,
it can be directly verified that $1/C_{\mathrm{III}}<3\pi/2$ and
so (\ref{eq:fesh2_cond}) implies $a<a_{H}=3\pi/4$ (see the proof
of Proposition~\ref{theo:fesh1}). We conclude that under the assumptions
of Proposition~\ref{theo:fesh2}, the assumptions of Proposition~\ref{theo:fesh1}
are fulfilled, too. Thus we have arrived at the estimates \[
\begin{split} & \|(H^{a}-\xi)^{-1}-(H_{\mathrm{eff}}^{a}-\xi)^{-1}\oplus0\|\leq\|(H^{a}-\xi)^{-1}-R_{\text{eff}}^{\mathscr{W}}(\xi)\oplus0\|\\
 & +\|R_{\mathrm{eff}}^{\mathscr{W}}(\xi)-(H_{\mathrm{eff}}^{a}-\xi)^{-1}\|\leq\left(\frac{8}{\pi}+\max\left\lbrace 1\,,\,\frac{2}{d_{\mathrm{eff}}(\xi)}\right\rbrace C_{\mathrm{III}}\right)\frac{2a}{d_{\text{eff}}(\xi)}+\frac{2a^{2}}{3\pi^{2}}\end{split}
\]
where we have used (\ref{eq:dist_ineq}). \end{proof}

Finally let us note that by combining Propositions~\ref{theo:coul_conv}
and \ref{theo:fesh3} one can show that a Hydrogen atom in a very
thin planar layer is well approximated, in the norm resolvent sense,
by the Coulomb-like potential in the plane.

\begin{theorem} \label{thm:Ha_HC_approx} Let $\eta\in\Res(H_{C})$
be such that $-3<\eta<0$, and $a>0$ fulfill \[
a<\min\left\lbrace a_{0},\,\frac{d_{C}(\eta)}{8C_{\mathrm{III}}}\right\rbrace \]
where $d_{C}(\eta)=\mathrm{dist}(\eta,\sigma(H_{C}))$ and $a_{0}$
is determined by the equation \[
\frac{2C_{\mathrm{I}}^{\,2}C_{\mathrm{II}}}{d_{C}(\eta)}\, a_{0}\,|\!\ln a_{0}|=\frac{1}{2}\,,\]
with $C_{\mathrm{I}}$ and $C_{\mathrm{II}}$ being defined in (\ref{eq:f_c_const})
and (\ref{eq:CII}), respectively. Then $\eta\in\Res(H^{a}-E_{1}^{a})$
and \begin{equation}
\|(H^{a}-E_{1}^{a}-\eta)^{-1}-(H_{C}-\eta)^{-1}\oplus0\|\leq\frac{4\, C_{\mathrm{I}}^{\,2}C_{\mathrm{II}}}{d_{C}(\eta)^{2}}\, a\,|\!\ln a|+\frac{20\, C_{\mathrm{III}}}{d_{C}(\eta)^{2}}\, a+\frac{2}{3\pi^{2}}\, a^{2},\label{eq:main_est}\end{equation}
with $C_{\mathrm{III}}$ being defined in (\ref{eq:c_fesh2}). \end{theorem}

\begin{proof} First apply Proposition~\ref{theo:coul_conv}, with
$\eta$ being substituted for $\xi$. By the above bound on $a$,
the assumptions of the proposition are fulfilled. Since $-3<\eta<0$
implies $d_{C}(\eta)<2$, it follows that $\eta\in\Res(H_{\mathrm{eff}}^{a}-E_{1}^{a})$
and \[
\|(H_{\mathrm{eff}}^{a}-E_{1}^{a}-\eta)^{-1}-(H_{C}-\eta)^{-1}\|\leq\frac{4\, C_{\mathrm{I}}^{\,2}C_{\mathrm{II}}}{d_{C}(\eta)^{2}}\, a\,|\!\ln a|.\]
According to (\ref{eq:dist_ineq_coul}), $d_{C}(\eta)\leq2d_{\mathrm{eff}}(E_{1}^{a}+\eta)$,
and this jointly with the choice of $\eta$ implies (\ref{eq:fesh2_cond})
for $\xi=E_{1}^{a}+\eta$. Hence the assumptions of Proposition~\ref{theo:fesh3}
are fulfilled, too. Thus $\eta\in\Res(H^{a}-E_{1}^{a})$ and we can
estimate \[
\begin{split} & \|(H^{a}-E_{1}^{a}-\eta)^{-1}-(H_{C}-\eta)^{-1}\oplus0\|\\
 & \leq\|(H^{a}-E_{1}^{a}-\eta)^{-1}-(H_{\mathrm{eff}}^{a}-E_{1}^{a}-\eta)^{-1}\oplus0\|+\|(H_{\mathrm{eff}}^{a}-E_{1}^{a}-\eta)^{-1}-(H_{C}-\eta)^{-1}\|\\
 & \leq\frac{4a}{d_{C}(\eta)}\left(\frac{8}{\pi}+\frac{4\, C_{\mathrm{III}}}{d_{C}(\eta)}\right)+\frac{2a^{2}}{3\pi^{2}}+\frac{4\, C_{\mathrm{I}}^{\,2}C_{\mathrm{II}}}{d_{C}(\eta)^{2}}\, a\,|\!\ln a|.\end{split}
\]
Observing that \[
\frac{8}{\pi}<\frac{C_{\mathrm{III}}}{2}<\frac{C_{\mathrm{III}}}{d_{C}(\eta)}\]
one arrives at (\ref{eq:main_est}). \end{proof}

Since the spectrum of $H_{C}$ is known explicitly one can use Theorem~\ref{thm:Ha_HC_approx}
to localize the point spectrum of the full Hamiltonian $H^{a}$ with
the aid of fairly standard perturbation methods \cite{kato}. We do
not pursue this problem here, however.

\section*{Acknowledgments}

The authors wish to acknowledge gratefully partial support from Grant
No. 201/09/0811 of the Czech Science Foundation and Grant No. LC06002
of the Ministry of Education of the Czech Republic. One of the authors
(M.T.) is also grateful for a partial support from Grant No. 202/08/H072
of the Czech Science Foundation.

\noindent \clearpage{}

\noindent \includegraphics{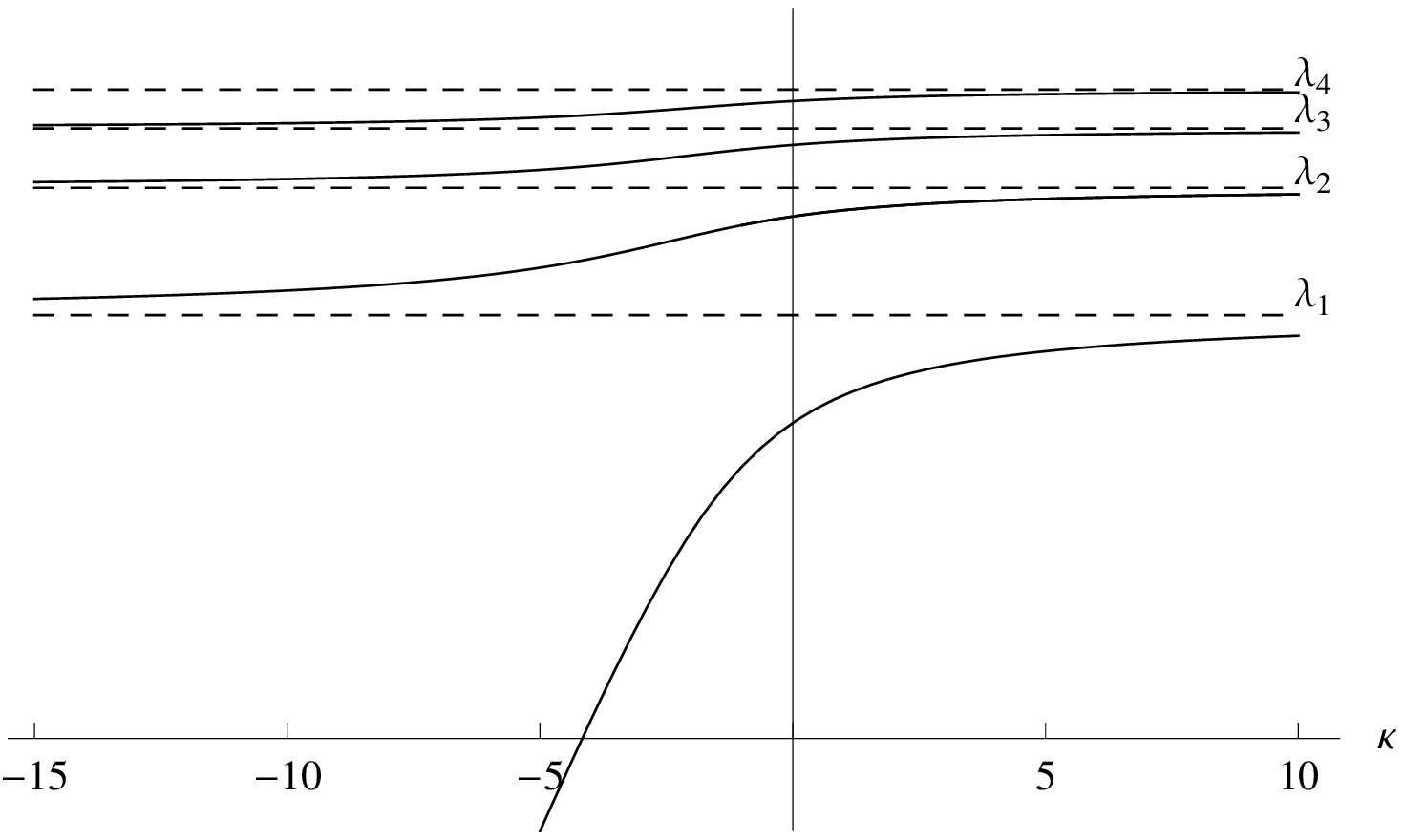}

\vspace{3\baselineskip}

\noindent Figure~1. Point levels of $H(\kappa)$. The vertical scale
is logarithmic.

\begin{thebibliography}{22}
\bibitem{as} M.~Abramowitz and I.~A. Stegun. \emph{Handbook of
Mathematical Functions}. (Dover, New York, 1970).

\bibitem{ag} S.~Albeverio, F.~Gesztesy, R.~H{\o}egh-Krohn and
H.~Holden (with an appendix by P.~Exner). \emph{Solvable Models
in Quantum Mechanics: Second Edition}. (AMS Chelsea Publishing, Providence,
2005).

\bibitem{AlMusallamTuan} F.~Al-Musallam and V.~K.~Tuan, A finite
and an infinite Whittaker integral transform. \emph{Computers Math.
Appl.} \textbf{46}, 1847--1859, 2003.

\bibitem{adl} O.~Atabek, C.~Deutsh and M.~Lavaud. Schr\"odinger
equation for two-dimensional Coulomb potential. \emph{Phys. Rev.~A}
\textbf{9}, 2617--2624, 1974.

\bibitem{BerezinShubin} F.~A.~Berezin and M.~A.~Shubin. \emph{The
Schr\"odinger Equation. }(Kluwer, Dordrecht, 1991).

\bibitem{bouzouina} A.~Bouzouina. Stability of the two-dimensional
Brown-Ravenhall operator. \emph{Proc. Roy. Soc. Edinburgh Sect.~A}
\textbf{132}, 1133--1144, 2002.

\bibitem{bd} R.~Brummelhuis and P.~Duclos. Effective Hamiltonians
for atoms in very strong magnetic fields. \emph{J.~Math. Phys.} \textbf{47},
art.~no. 032103, 2006.

\bibitem{bgl} J.~Br\"uning, V.~Geyler and I.~Lobanov. Spectral
properties of a short-range impurity in a quantum dot. \emph{J.~Math.
Phys.} \textbf{46}, 1267--1290, 2004.

\bibitem{bulla_ges} W.~Bulla and F.~Gesztesy. Deficiency indices
and singular boundary conditions in quantum mechanics. \emph{J.~Math.
Phys.} \textbf{26}, 2520--2528, 1985.

\bibitem{dov} C.~R.~de~Oliveira and A.~A.~Verri. Self-adjoint
extensions of Coulomb systems in 1,2 and 3 dimensions. \emph{Annals
of Physics} \textbf{324}, 251--266, 2009.

\bibitem{GesztesyZinchenko} F.~Gesztesy and M.~Zinchenko. On spectral
theory for Schr\"odinger operators with strongly singular potentials,
\emph{Math. Nachr.} \textbf{279}, 1041--1082, 2006.

\bibitem{herbst} I.~W.~Herbst. Spectral theory of the operator
$(p^{2}+m^{2})^{1/2}-Ze^{2}/r$. \emph{Commun. Math. Phys.} \textbf{53},
285--294, 1977.

\bibitem{hostler} L.~C.~Hostler. Coulomb Green's function in $f$-dimensional
space. \emph{J.~Math. Phys.} \textbf{11}, 2966--2970, 1970.

\bibitem{Fulton} Ch.~Fulton. Titchmarsh-Weyl $m$-functions for
second-order Sturm-Liouville problems with two singular endpoints.
\emph{Math. Nachr.} \textbf{281}, 1418--1475, 2008.

\bibitem{kato} T.~Kato. \emph{Perturbation theory for linear operators}.
(Springer-Verlag, New York, 1966).

\bibitem{pp} D.~G.~W.~Parfitt and M.~E.~Portnoi. The two-dimensional
hydrogen atom revisited. \emph{J.~Math. Phys.} \textbf{43}, 4681--4691,
2002.

\bibitem{rs2} M.~Reed and B.~Simon. \emph{Methods of Modern Mathematical
Physics II}. (Academic Press, New York, 1975).

\bibitem{rs3} M.~Reed and B.~Simon. \emph{Methods of Modern Mathematical
Physics III}. (Academic Press, New York, 1979).

\bibitem{rs4} M.~Reed and B.~Simon. \emph{Methods of Modern Mathematical
Physics IV}. (Academic Press, New York, 1978).

\bibitem{weidmann} J.~Weidmann. \emph{Linear Operators in Hilbert
Spaces}. (Springer-Verlag, New York, 1980).

\bibitem{ygc} X.~L.~Yang, S.~H.~Guo and F.~T.~Chan. Analytic
solution of a two-dimensional hydrogen atom. I. Nonrelativistic theory.
\emph{Phys. Rev.~A} \textbf{43}, 1186--1196, 1991.

\bibitem{WangZhu} Z.-Q.~Wang and M.~Zhu. Hardy inequalities with
boundary terms. Electronic J.~Diff. Equations \textbf{43}, 1--8,
2003.

\end{thebibliography}
\end{document}